\documentclass[
reprint,
superscriptaddress,
 amsmath,amssymb,
 aps,
pra
]{revtex4-2}
\usepackage{graphicx} 
\usepackage{xcolor}
\usepackage{siunitx}
\usepackage{hyperref}
\usepackage{ragged2e}
\usepackage{gensymb}
\begin{document}

\title{Permanent Magnet Electron Optics for Low Energy Electron Systems:\\The Art of Extraordinary Performance from Ordinary Components}

\author{Ameya Patwardhan}
    \affiliation{Department of Applied Physics and Science Education\\
    Eindhoven University of Technology, P.O. Box 513, 5600 MB Eindhoven, The Netherlands}
\author{Bas van der Geer}
    \affiliation{Pulsar Physics, 5614 BC Eindhoven, The Netherlands}
\author{Jom Luiten}
    \affiliation{Department of Applied Physics and Science Education\\
    Eindhoven University of Technology, P.O. Box 513, 5600 MB Eindhoven, The Netherlands}
\author{Julius Huijts}
    \email{j.v.huijts@tue.nl}
    \affiliation{Department of Applied Physics and Science Education\\
    Eindhoven University of Technology, P.O. Box 513, 5600 MB Eindhoven, The Netherlands}
\date{\today}

\begin{abstract}
Permanent magnet electron optics offer many advantages over electromagnets, and are being increasingly used in high energy (\si{\giga\electronvolt}) electron accelerator designs. 
Here, we identify the advantages of permanent magnet electron optics for low energy (\si{\kilo\electronvolt}) electron accelerators.
We explore the applications of a class of designs based on axially magnetized permanent magnets, which offer a variety of advantages such as short focal lengths (few \si{\milli\meter}), while also preventing apparent emittance growth resulting from starting particles in a magnetic field. 
The proposed design philosophy is applied to an accelerator based on the ultracold electron source. 
The design is shown to be `emittance preserving' even for very short focal lengths ($\sim$\SI{5}{\milli\meter}) at an emittance level better than \SI{1}{\nano\meter\radian}, while the short beamline ($\approx$ \SI{12}{\centi\meter}) limits space-charge effects. 
Two remedies for the mitigation of typical manufacturing and alignment challenges are considered. 
The performance of the design (related to parasitic aberrations) is enhanced by the proposed techniques. 
Applications of this design philosophy can improve the performance of ultrafast electron diffraction setups with minimal manufacturing effort.  
\end{abstract}

\maketitle

\section{Introduction}
\label{sec:intro}
The use of permanent magnets instead of electromagnets is being increasingly explored in the accelerator community for reasons of compactness, energy efficiency and absence of noise.  
While permanent magnets have routinely been used in the past for undulators, increasingly accelerator designs have included permanent magnet dipoles such as the Swiss Light Source upgrade project\cite{willmott_sls_2024, sanfilippo_magnets_2024}.

In addition to these radiation sources, permanent magnets are also increasingly employed to control electron beam dynamics, such as the tunable permanent magnet based quadrupole\cite{halbach1980design, lou1998stability, dong_development_2025, barna_tunable_2025}.

In this article, we explore the use of permanent magnets for low-energy (\si{\kilo\electronvolt}) electron beamlines. We consider the combination of a permanent magnet focusing system with a DC accelerator, making our findings relevant to e.g. UED setups \cite{siwick2003atomic,vanoudheusden2010compression}, DC-RF accelerators \cite{rao_engineering_2014} and systems for electron microscopy \cite{egerton2005physical} and lithography\cite{chen2015nanofabrication}. Our specific design case is the development of a microfocus beamline based on an ultracold (or cold-atom-based) electron source\cite{claessens_cold_2007,luiten2007ultracold,franssen_compact_2019}.


\section{Motivation for using Permanent Magnets: High Performance System Design}
\label{sec:motivation}
For low energy electrons it may not be immediately clear why it would be desirable to use permanent magnets, since relatively low power is needed for electromagnets. The motivation can be derived from two challenging aspects of electromagnets.
The first concerns the stability and noise of power supplies. Noise and ripple on electromagnet current effectively degrade instrument performance, particularly relevant for low-emittance beams, where they can for example limit resolution in electron microscopy and lithography.
While the high frequency noise can be damped by suitable filters, ensuring low drift is challenging.

The second aspect is mechanical integration.
In the aforementioned applications, it is desirable to ensure a small longitudinal spread of the magnetic fields to ensure that a complex `column' can be constructed in a limited length.
In order to achieve this goal, the electromagnets need to be placed in vacuum, which poses constraints on the materials and methods of construction. 
In addition, the electromagnet must now necessarily be cooled (as opposed to possibly passive cooling outside of vacuum) which further complicates the mechanical design. 

Permanent magnets, in contrast, are practically noise-free, are only affected by negligible long-term drift (typically ppm per year \cite{zingery_evaluation_1966}) and do not require cooling in vacuum, allowing more compact, simple designs. 

But, as we consider the combination of a magnetic focusing system with a DC accelerator, another question comes into play: the alignment between the two. As we will show in Section (\ref{sec:embedded}), using permanent magnets allows for a compact, `self-aligned' solution that is not possible with electromagnets.

\section{Apparent Emittance}
\label{sec:appemittgrowth}
A high-brightness/low-emittance electron beam setup almost always consists of some magnetic focusing element for the purpose of focusing the beam onto the sample/detector/next accelerator element. 
Typically, for non-relativistic electron beam systems (or photoinjectors), solenoids are used as the focusing element.
In practice, the magnetic field of a solenoid extends beyond its geometric limits, creating a non-zero axial magnetic field at the source and focus. 
The interaction between the electron beam and this axial fringe field causes apparent emittance growth, which is undesirable for low-emittance beams. 
This apparent emittance growth is related to the `angular momentum' of the magnetic field. 
The apparent (normalized rms) emittance due to non-zero axial magnetic field ($B_z$) can be expressed as \cite{rao_engineering_2014, reiser_theory_2008} 
\begin{equation}
    \label{eq:apparent_emittance}
    \epsilon_{n} = \frac{e B_z r_0^2}{8 m c}
\end{equation}
where $r_0$ is the radius of the beam ($r_0$ defined as $\sqrt{\sigma_x^2 + \sigma_y^2}$). 
Therefore, in applications where beam emittance is critical, a `bucking solenoid' is often used behind the source, which counteracts the magnetic field of downstream solenoids to produce a zero axial magnetic field at the source.
In the context of ultrashort pulsed beams and/or large bunch charges, the source size can be $\sim$ \SI{1}{\milli\meter} to reduce space-charge effects and in such a case, it is important to ensure the zero of the magnetic field.
Alternatively, appropriate shielding and large distances can be used to minimize the magnetic field at source and focus, but this is not always possible, especially for compact injectors and UED setups where propagation at subrelativistic energies is kept at a minimum to prevent longitudinal expansion of the bunch as well as to limit space-charge related emittance growth.
It is also clear from Equation (\ref{eq:apparent_emittance}), that for best performance with low-emittance beams it is desirable to have the magnetic field be zero at the focus as well. 
Electron microscopy or diffraction studies on magnetic samples may also be a reason to require zero magnetic field at the focus\cite{shibata_atomic_2019, tsuno_magnetic-field-free_1983, kohno_new_2017, petford2005lorentz}.  
In the next section, we present a magnet design that ensures presence of two magnetic zeros.  

\section{Axially Magnetized Magnets}
\label{sec:magnet}
\begin{figure}
    \centering
    \includegraphics[width=\linewidth]{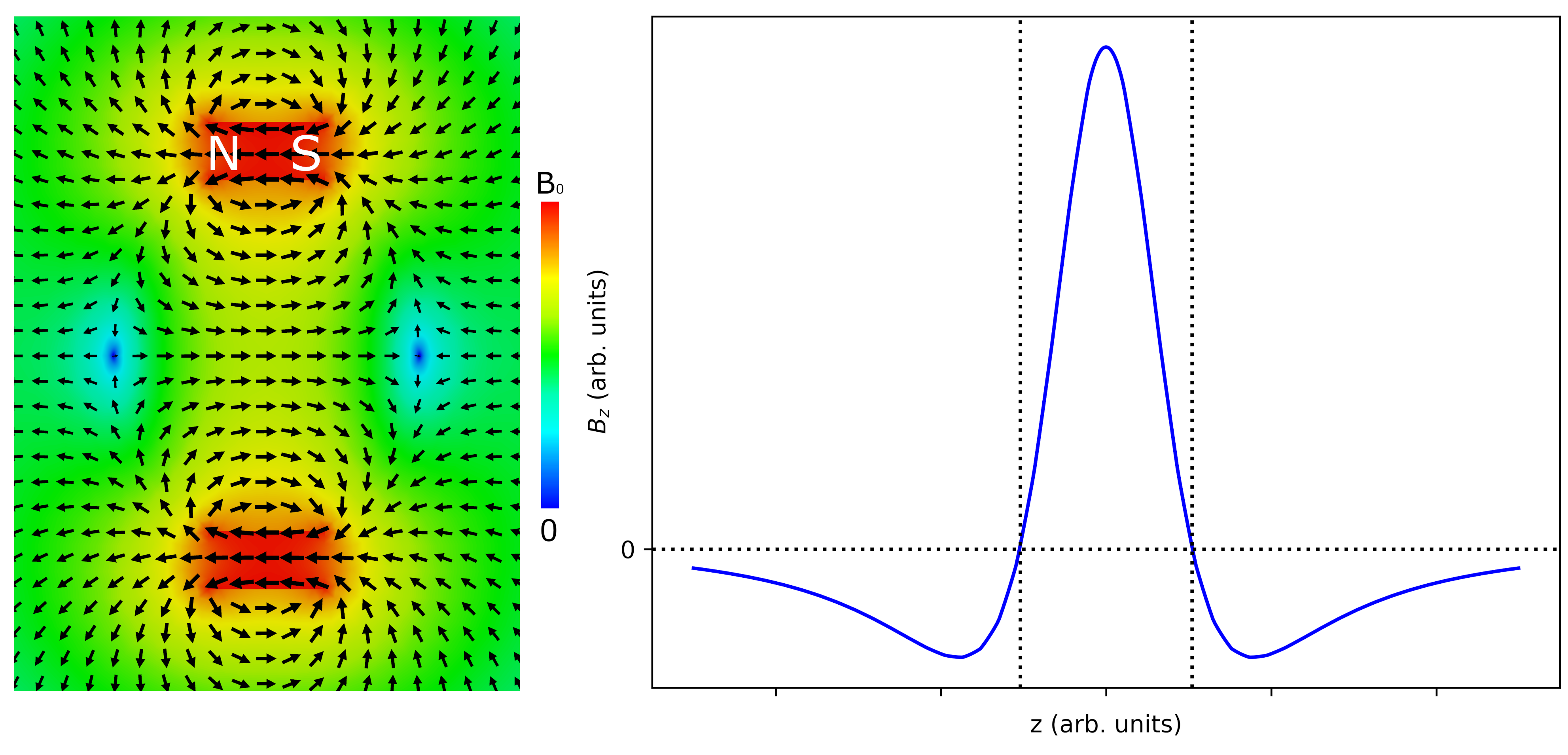}
    \caption{\textbf{Field of an Axially Magnetized Ring Shaped Permanent Magnet.} \textit{Left:} The field lines are plotted on a cross-section of the magnet through the symmetry axis. \textit{Right:} The plot shows the axial magnetic field $B_z$ as a function of axial coordinate $z$. The two zeros of the magnetic field are indicated at the intersection of dashed lines.}
    \label{fig:PermanentMagnet}
\end{figure}
While it is possible to achieve these zero-field points by adding bucking solenoids for both the source and focus point, we propose that the same can be achieved with a single axially magnetized ring-shaped permanent magnet.
The field lines of such a permanent magnet geometry are shown in Figure (\ref{fig:PermanentMagnet}). Although such a magnet geometry is already known \cite{jayamanna_single_2018, gehrke_design_2013}, to the best knowledge of the authors, its use in preventing apparent emittance growth has not been explored.
The axial magnetization of such a ring-shaped magnet ensures the presence of two such zero points of magnetic field.
A convenient feature of this configuration is that the shape of the magnetic field, and therefore the location of the zeros, depends only on the magnet geometry, and not on the exact value of the magnetization. 
This is extremely useful because the magnetization process has a typical tolerance in the range of 3 - 5 \%, which will thus only provide a scaling to the field, leaving the position of the zeros unaffected.  
From the preceding discussion, it follows that the magnet should be constructed such that the first zero coincides with the source, and the second coincides with the focus, possibly respecting additional constraints.
Next, we introduce an optimization strategy based on genetic optimization to solve the constraint problem.

\section{Optimization Strategy}
\label{sec:optimization}
We first discuss some constraints on the magnetic fields.
Although the representative schematic of the magnetic field as shown in Figure (\ref{fig:PermanentMagnet}) has mirror symmetry in $z$, it may be desirable to achieve an asymmetrical configuration for reasons stated below.
Accelerating the electrons between the source and the focus makes the beam dynamics asymmetrical, irrespective of the symmetry of the magnetic field.
From the beam-dynamics perspective, it may be desirable to have a specific magnetic field gradient at the source to compensate the transverse blow-up of the beam due to space-charge effects or for emittance compensation of RF photo-injectors.
Additionally, the shape and magnetization of the magnet (and possibly the acceleration gradient and length) can be varied in order to achieve the optimal configuration.

In practice, instead of machining a single magnet to a specific shape and accurately controlling its magnetization, it is more practical to approach the ideal shape and magnetization by appropriate tiling with smaller magnets.
By adopting an appropriate packing fraction, the field profile can be fine-tuned. 
Although such a tiled arrangement is contrary to the guaranteed location of the zeros mentioned above, in practice it allows a higher degree of control on the field profile.
As stated above, the challenge is to find a magnet configuration that optimizes the quantities of interest subject to various constraints on the magnetic field as well as any other constraints of the beamline.

We choose genetic optimization as a suitable optimization strategy because of its ability to explore complete parameter spaces without getting stuck at a local minimum.
More formally, a set of constraints for the genetic optimization consist of: 
\begin{itemize}
    \item Magnetic field $B_z|_\textnormal{z=source} \Rightarrow 0$ 
    \item Magnetic field $B_z|_\textnormal{z=focus} \Rightarrow 0$
    \item Magnetic field gradient $\dfrac{\partial B_z}{\partial z}|_\textnormal{z=source} \Rightarrow B_0^{'}$
    \item Any geometric constraints on the location of the magnet(s) along the beamline axis.
\end{itemize}

A set of optimization criteria consist of: 
\begin{itemize}
    \item Minimize beam radius $r|_\textnormal{z=focus}$
    \item Maximize charge $Q|_\textnormal{z=focus}$
\end{itemize}

A set of optimization variables consists of: 
\begin{itemize}
    \item Geometric properties of the magnet(s)
    \item Properties of the source
\end{itemize}

As the problem has more than one optimization criterion it is a `multi-objective optimization problem', while the set of optimized (`Pareto-efficient') solutions forms a Pareto front (see e.g. \cite{DebMOO2005}). This Pareto front is the outcome of the optimization procedure, showing the trade-off between the two optimization criteria. Based on this Pareto front, an optimally-informed design choice can be made. 
We will now use this scheme to design a micro-focus beamline for a high-brightness electron injector based on the ultracold electron source. 
This application is especially demanding because it requires the shortest focal lengths and thus requires the highest magnetic fields, while still respecting the constraints stated above. 

\section{The Ultra-Cold Electron Source}
\label{sec:uces}
The ultracold electron source was proposed in \cite{claessens_ultracold_2005} and realized in subsequent versions in Eindhoven \cite{claessens_cold_2007,taban2010ultracold,engelen2013high-coherence,franssen_compact_2019} and Melbourne \cite{mcculloch2013high}. 
It is based on the photo-ionization of laser-cooled Rubidium atoms to produce an electron bunch with a record-low bunch temperature on the order of \SI{10}{K} - i.e. a mean transverse energy (MTE) on the order of \SI{1}{meV}. 
Uniquely, the electrons are generated from a three-dimensional `cathode' (the ionizing fraction of the laser-cooled atoms), the shape of which can be controlled in 3D. 
This in turn allows shaping of the initial phase space distribution of the electron beam\cite{mcculloch2011arbitrarily}. Also, each shot is generated from a `fresh' cathode of neutral atoms.
The generated electron bunches can have sub-picosecond duration \cite{de_raadt_subpicosecond_2023, de_raadt_ultracold_2024}, but bunch charge has so far been limited to the order of \si{fC}. 
Finally, the use of laser-cooled atoms introduces two requirements that are relevant to the present design. First, the laser-cooled atoms are trapped in the center of a magneto-optical trap (MOT), which needs to be a quadrupolar zero of the magnetic field with a certain magnetic field gradient (see e.g. \cite{metcalf1999laser}).
Second, the DC electric field used to accelerate the electron beam imparts Stark shifts on the different atomic levels used in laser cooling. For the laser-cooling to be efficient, the accelerating field is limited to about about \SI{1}{\mega\volt/\meter} \cite{krenn_stark_1997, franssen_ultracold_dissertation_2019}.

\section{Microfocus Beamline design based on an ultracold electron source}\label{sec:design}
\begin{figure}
    \centering
    \includegraphics[width=0.85\linewidth]{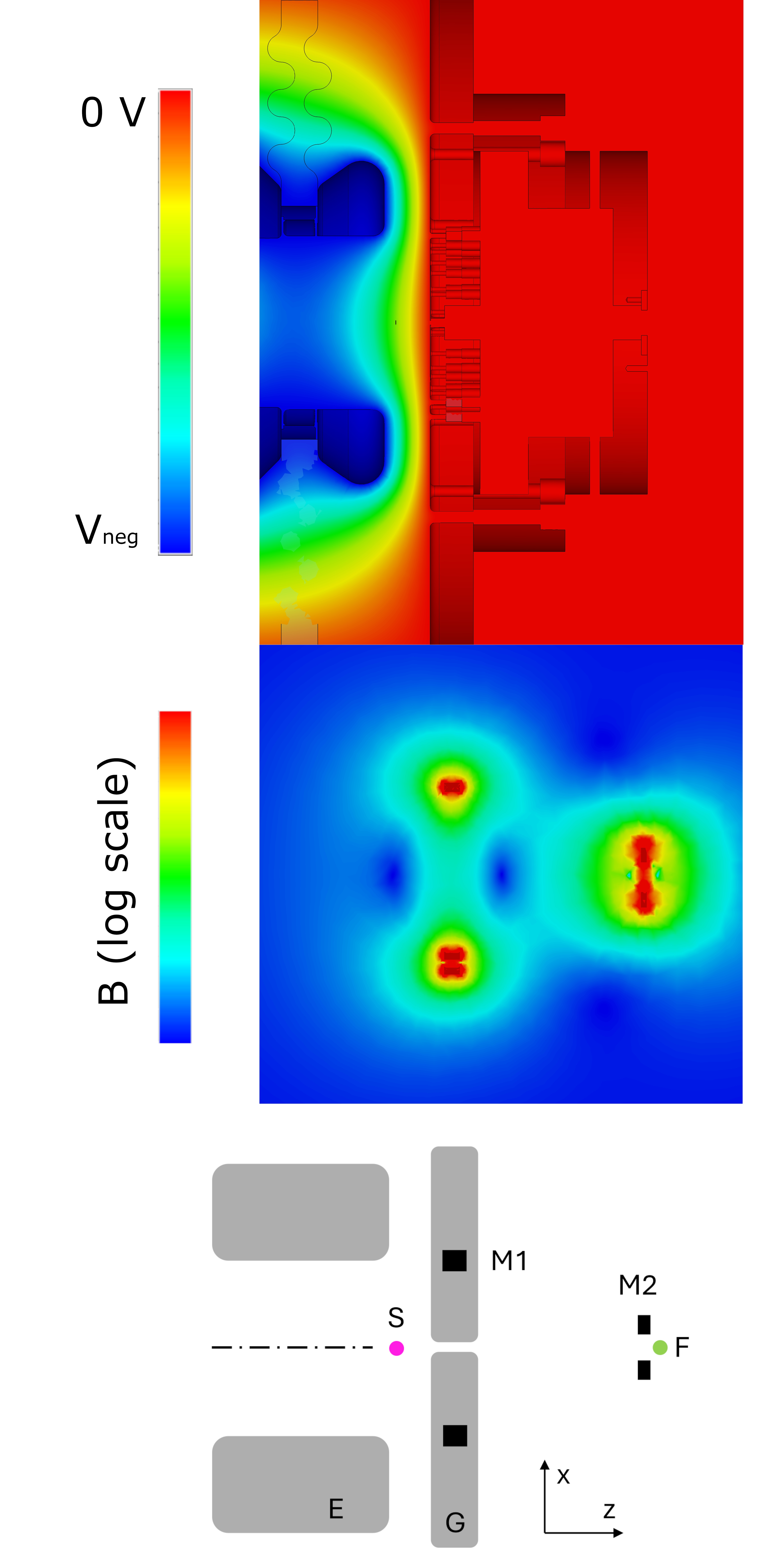}
    \caption{\textbf{Electromagnetic design of the accelerator and magnet assembly.} \textit{Top:} Electrostatic Potential: The color scheme represents the electric potential, with red representing ground and blue representing the highest negative potential. \textit{Center:} The magnetic field strength in the same structure in a section passing through the symmetry axis. The color scheme is chosen to be logarithmic to illustrate the zero-field points. The discretization causes some loss to the symmetry of the field in the immediate vicinity of the magnet (M1), but this asymmetry is averaged out close to the symmetry axis. \textit{Bottom:} Basic schematic of the same structure, ground electrodes (G) and the negative electrode (E) in grey and magnets (M1 and M2) in black. The source (S) is illustrated as a pink dot and the focal point (F) as a green dot. The origin of the coordinate system is at the centroid of the source, with the z-axis corresponding to the symmetry axis of the accelerator.}
    \label{fig:RbDCAccelerator}
\end{figure}

\begin{figure}
    \centering
    \includegraphics[width=0.8\linewidth]{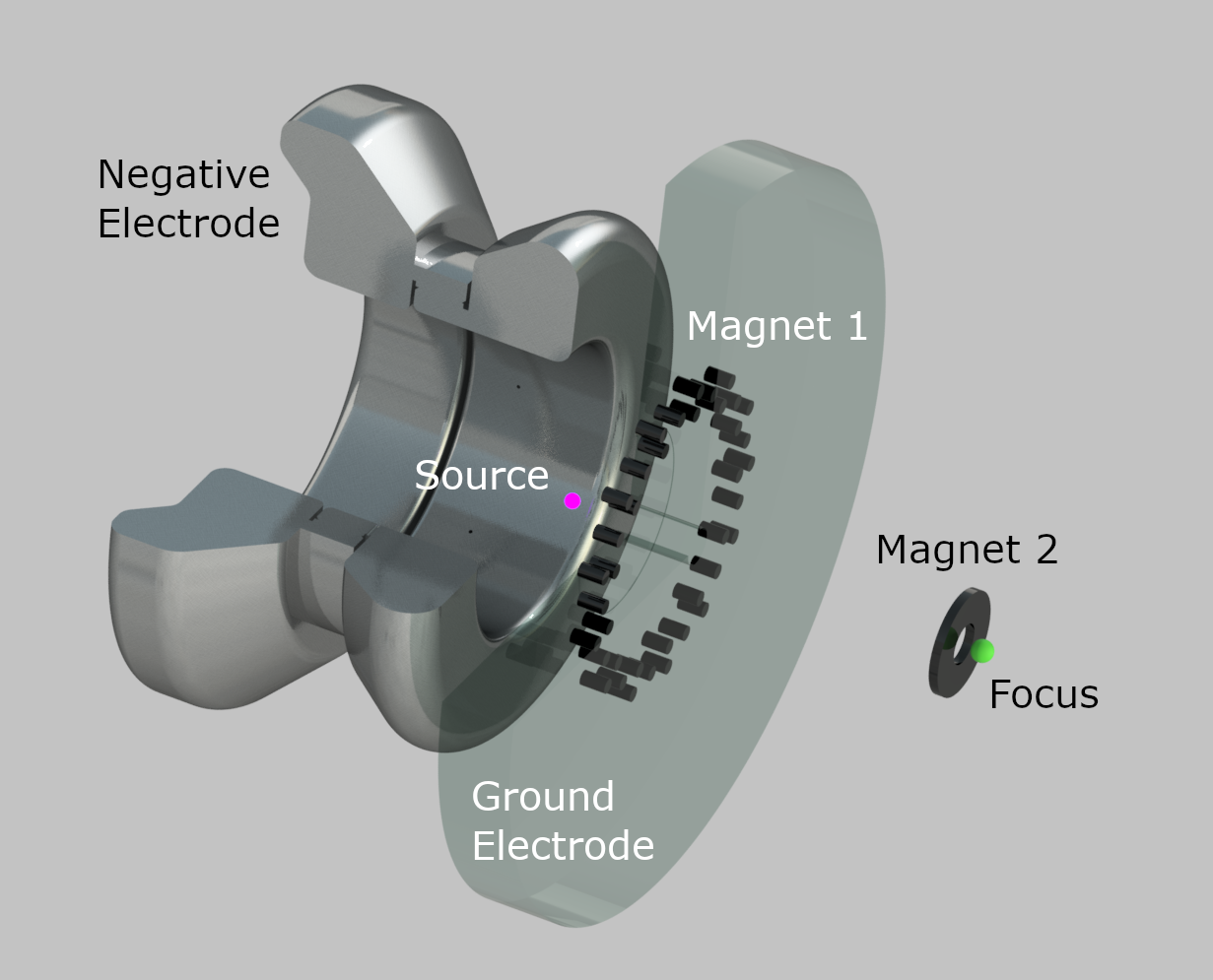}
    \caption{\textbf{Illustration of the proposed design based on the ultracold electron source.} The pink dot represents the laser cooled Rb atom cloud. The first `magnet' is discretized in a symmetrically patterned array of axially magnetized magnets. The second magnet is ring shaped with a rectangular cross-section. The green dot represents the focal point of the electron beam.}
    \label{fig:FullAcceleratorDesign}
\end{figure}

At a generic level, the accelerator design consists of a DC accelerator assembly and a focusing assembly. The features and designs of both are highlighted below. A representative schematic of the accelerator and magnet assembly is shown in Figure (\ref{fig:RbDCAccelerator}), alongside electrostatic potential and magnetic field maps illustrating the zeros of the magnetic field. A simplified CAD drawing of the combined assembly is shown in Figure (\ref{fig:FullAcceleratorDesign}).

\subsection{DC Accelerator}
The DC accelerator design of the ultracold electron source is heavily influenced by the requirements of atom trapping mentioned above. 
Furthermore, like our current ultracold electron source (\cite{franssen_compact_2019}), the proposed injector employs the grating magneto-optical trap \cite{cotter_design_2016, mcgilligan_grating_2017, nshii_surface-patterned_2013}, which necessitates on-axis optical access to the DC accelerator. 
While in existing designs, this is achieved through an indium tin oxide coated window as the negative electrode, this introduces issues related to charging of the window \cite{de_raadt_ultracold_2024}, which we overcome in this new design through the use of a cylindrical electrode with a central bore (`E' in Figure (\ref{fig:RbDCAccelerator})) to enable optical access of the trapping laser. 
The design requires a nominal operating voltage of \SI{-40.6}{\kilo\volt} for the electrode in order to produce a \SI{16.6}{\kilo\electronvolt} beam.
This is because the Rb atoms are situated in between the negative electrode and ground electrode (`S' in Figure (\ref{fig:RbDCAccelerator})).

\subsection{Magnet Assembly}
As introduced above, the atom trapping occurs at a point of zero magnetic field, with a region of quadrupolar magnetic field around the zero point. 
The quadrupolar nature of the zero can include rotational symmetry, i.e., the quadrupolar symmetry can be in the x-z plane while allowing rotational symmetry around the z-axis.
This coincides with the requirement introduced in Section (\ref{sec:appemittgrowth}): to minimize apparent emittance growth, the magnetic field should be zero at the source. 
While in particle accelerators, the optimal value of the magnetic field gradient at the source is set by space-charge effects and beam dynamics, for atom trapping the optimal gradient is related to the trap size and trap density required. 
The requirement for atom-trapping supersedes any optimality conditions from the beam dynamics perspective. 
In the case of Rb atoms, a suitable value for the magnetic field gradient is between \SIrange[range-phrase=\,and\,]{0.1}{0.2}{\tesla / \meter} \cite{franssen_ultracold_dissertation_2019, yoon_characteristics_2007}.
Geometric constraints were also placed to ensure that the magnets were placed downstream of the ground electrode and at least \SI{5}{\milli\meter} upstream of the `focal' point to ensure mechanical access for probes/diagnostics at the focus. 

To test the limits of this design methodology and of the magnet design, it is desirable to have as small a focal spot as possible while maximizing the bunch charge within the performance limits of the source. 
The magnet optimization process involves varying the parametrized geometrical cross-section of the magnet, as well as the tiling of multiple small magnets as introduced above.
Additionally, the source shape and volume was allowed to be varied, with minimum and maximum extents of \SI{30}{\micro\meter} and \SI{100}{\micro\meter} for both transverse and longitudinal dimensions. 
This additional degree of freedom, not present in conventional electron sources, allows for an optimal tradeoff between transverse emittance, longitudinal emittance and bunch charge.

In line with the design principles stated above, the magnet design was optimized with genetic optimization as built into the General Particle Tracer\cite{pulsar_physics_general_nodate}.
The optimization process was first tried out with a single magnet (similar to the one shown in in Figure (\ref{fig:PermanentMagnet})), though with a cross-section that was allowed to be non-orthogonal.
The optimal magnet configuration produced a focal spot size of approximately $\sigma_r=$ \SI{2.5}{\micro\meter}.
In the view of the authors, the geometric structure of the magnet was considered to be difficult to realize in practice.
Consequently, a solution space consisting of two magnets was explored.
This enables a sub-micron focal spot size. 
This magnet configuration is shown in Figure (\ref{fig:FullAcceleratorDesign}).
The optimal solution consists of a larger magnet that generates the necessary magnetic field for the magneto-optical trapping, but due to its larger radial extent, it also contributes to some extent to the focusing of the electron beam.
The smaller magnet is mainly the focusing element with a short focal length and is placed very close to the focal point.
It is convenient to tile the larger magnet because comparable performance can be achieved using off the shelf magnets. 
The resulting magnetic field error when compared to the `optimal' solution is about 1 \% with a 4 \% increase to the focal spot size, which is an acceptable compromise.
Finally, we allow tuning of the accelerating potential to accommodate a 10 \% error margin in the strength of the magnets, exceeding typical manufacturing tolerances of magnets.

\begin{figure}
    \centering
    \includegraphics[width=0.95\linewidth]{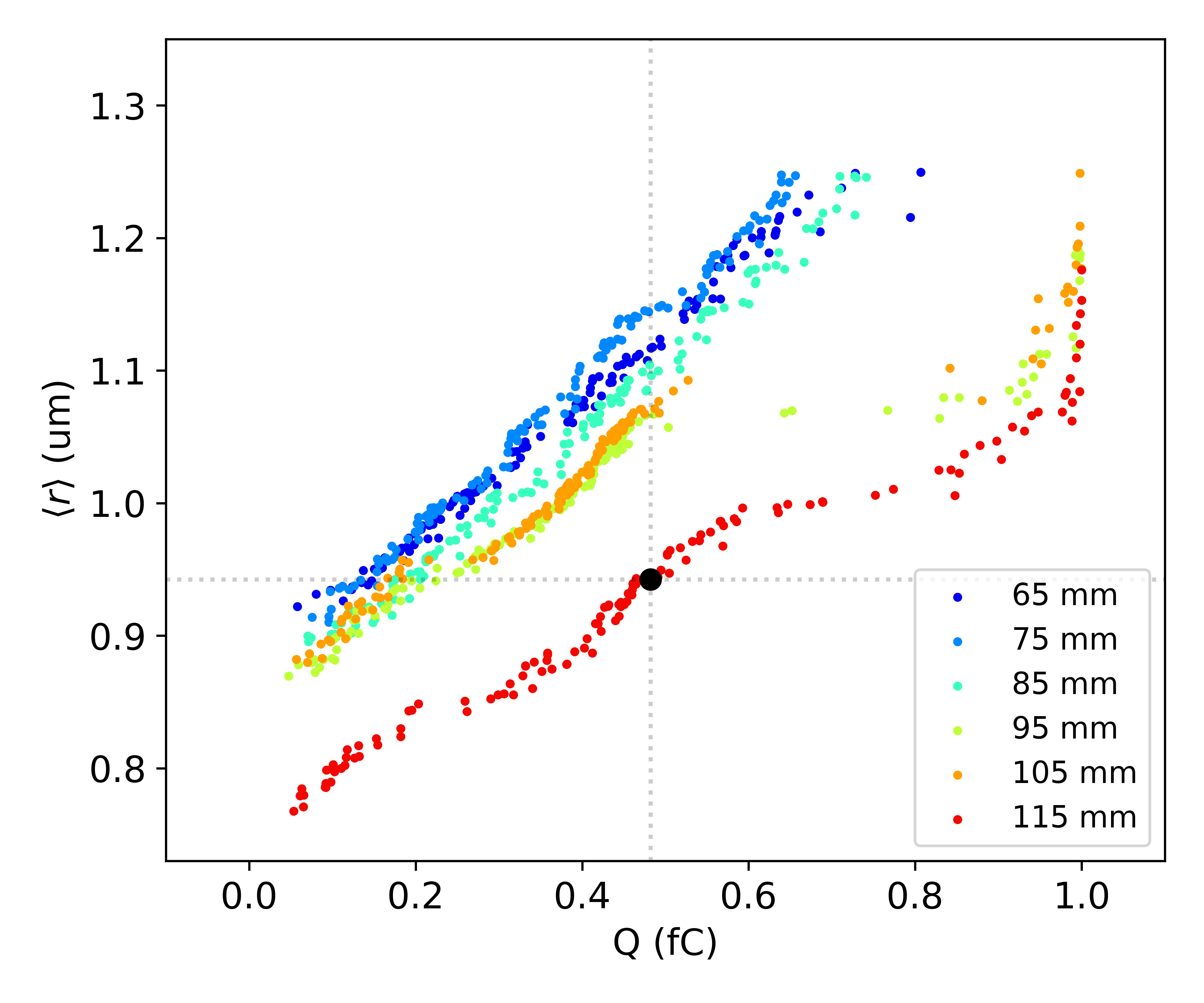}
    \caption{\textbf{Pareto fronts for different focal positions:} Each Pareto front shows the trade-off between the two optimization criteria, i.e. focal spot size $\left<r\right>$ ($r = \sqrt{x^2 + y^2}$) and bunch charge $Q$. The black dot indicates the selected design.}
    \label{fig:ParetoFront_f}
\end{figure}

The optimization process was carried out for six different positions of the focus, thus producing six Pareto fronts, shown in Figure (\ref{fig:ParetoFront_f}). 
Each point in the graph corresponds to a different magnet geometry.
Each Pareto front shows the trade-off between the two optimization criteria, i.e. the focal spot size and the bunch charge.
Clearly, moving the focus further downstream reduces the focal spot size, which is because the beam is slightly diverging before the focusing magnet (Magnet 2).
Moving the focus downstream also moves Magnet 2 downstream, increasing the beam size at the magnet thus causing a tighter focus.
We limit the focal position to 115 mm to keep the design compact and limit pointing instabilities. 
Since this limit offers the smallest focal spot size we select a design from this Pareto front.
For this, we choose a bunch charge of \qty{0.5}{fC}, a conservative value readily achieved with our current ultracold electron source.
A reasonable operating regime of this design in terms of space-charge effects is described in the supplementary information.

\subsection{Particle Tracking Simulations}\label{sec:results}
\begin{figure}
    \centering
    \includegraphics[width=0.95\linewidth]{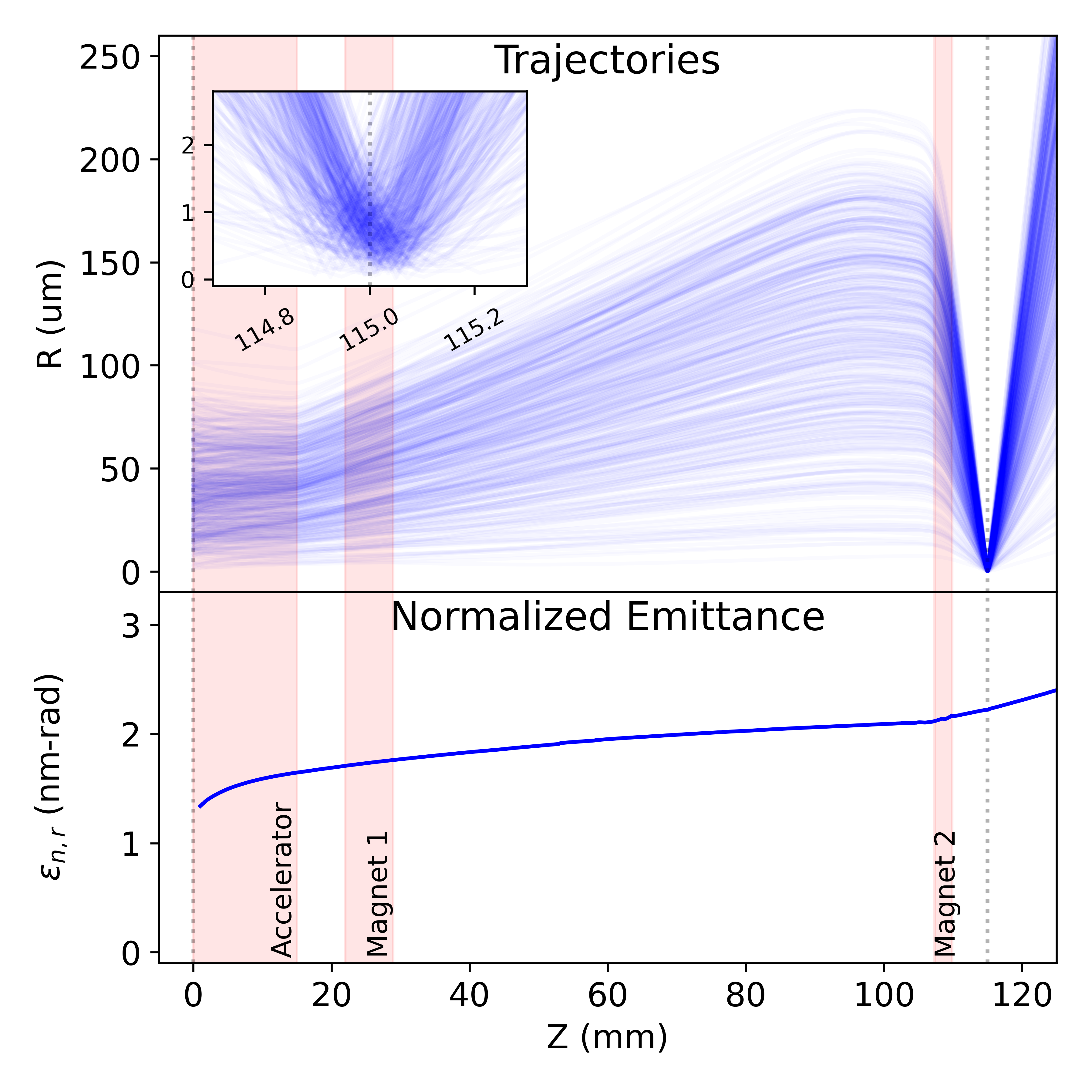}
    \caption{\textbf{Particle Tracking Results:} \textit{Top:} The trajectories of a random sample of the electrons are shown. The inset shows a smaller portion of the trajectories near the focal point. \textit{Bottom:} The normalized emittance as calculated by the General Particle Tracer nemirrms routine is shown.}
    \label{fig:ParticleTracking}
\end{figure}
The particle tracking simulations were performed using the General Particle Tracer with fieldmaps generated by CST, Traceon (now Voltrace) and custom elements \cite{pulsar_physics_general_nodate, noauthor_voltrace_nodate, noauthor_cst_nodate}. 
Using multiple `field-maps' provides a sanity check against suboptimal meshing or convergence settings.
The plots in Figure (\ref{fig:ParticleTracking}) show the results of beam dynamics simulations.
The trajectories show a region of acceleration (from \SI{0}{\milli\meter} to \SI{15}{\milli\meter}) where the beam becomes divergent due to the negative lens effect of the ground electrode. 
The beam continues to diverge until the second magnet is encountered, which creates a tight focus. 
The slow emittance growth is attributed to space-charge effects upto the second magnet ($z\approx$ \SI{105}{\milli\meter}).
Beyond this, the emittance growth can be attributed to misaligned slices in the transverse phase-space i.e. chromatic aberration.
See supplementary information (Figure (\ref{fig:sc_chromatic})). 

While typically space-charge effects are dominant at a focus, in our case the focus is so tight that the pulse length exceeds the Rayleigh length, i.e. at any given moment only a fraction of the bunch is tightly focused. As a result, emittance growth at the focus is negligible.
A focal spot size of $\sigma_r=$ \SI{920}{\nano\meter} ($\left<r\right>$=\SI{780}{\nano\meter}) is obtained for a beam energy of \SI{16.6}{\kilo\electronvolt} and a bunch charge of \SI{0.5}{\femto\coulomb}.
The energy spread of the beam is 0.2 \% ($\sigma_E/E$), with a bunch length of $\sigma_t= $\SI{1.4}{\pico\second}.
While the divergence angle of the beam at the focus is relatively large ($\sigma_\theta=$ \SI{18.8}{\milli\radian}), the Rayleigh length is in the order of \SI{100}{\micro\meter}. 
The robustness of the design in terms of the focal spot size and the effects of apparent emittance growth are quantified in the supplementary information.

\section{Self-Aligned Magnetic Field: Magnets embedded inside a high voltage electrode}
\label{sec:embedded}
The preceding section tacitly assumes that the DC accelerator and the magnets are aligned to one another exactly, which cannot be the case in practice. This alignment concerns both the offset and the tilt of the axes with respect to each other. 

In the case of the DC high voltage assemblies, the challenge of alignment is amplified due to the necessity of using an insulator as a mechanical support. 
It is important to note that the symmetry axes of a DC accelerator are actually two separate datums, the datum of the cathode and the datum of the anode (or ground). 
These datums are necessarily misaligned, and the misalignment is amplified due to the large creep distance (and thus large dimension) of the high voltage insulator.
Aligning a magnet to the ground electrode is simple because anything placed upstream of the anode is automatically grounded and shielded. 
However, this ultimately increases the beamline length.
In the case of ultrafast electron diffraction, it is desirable to limit the beamline to the smallest possible dimension to reduce bunch length.
In the case of DC-RF Photoinjectors, it may also be desirable to separate the magnetic field from the RF cavity for example to prevent multipacting. 
For both these situations, placing the magnet inside the high voltage electrode can enable a compact design that is self aligned. 
By self-alignment, we refer to the idea that the magnets (and hence the magnetic field) are mechanically referenced directly to the high voltage electrode without intermediate `parts' that need to transfer the accuracy of the reference features.

Embedding magnets inside a high voltage electrode can be achieved with relative ease, with care being needed to ensure that no magnet feature has a high electric field gradient; a challenge which can be addressed by recessing the magnets with respect to an outward surface of the electrode.
Note that achieving a similar result with electromagnets requires floating a high current supply ($\approx$\SI{10}{\ampere} and $\approx$ \SI{100}{\watt}) at high voltages ($\approx$ \qtyrange[range-units=single]{1}{100}{\kilo\electronvolt}), in addition to e.g. galvanic isolation of the cooling circuit. 
In contrast, a permanent magnet needs no floating high current supply and no cooling with the added benefit of low drift.

\begin{figure}
    \centering
    \includegraphics[width=0.85\linewidth]{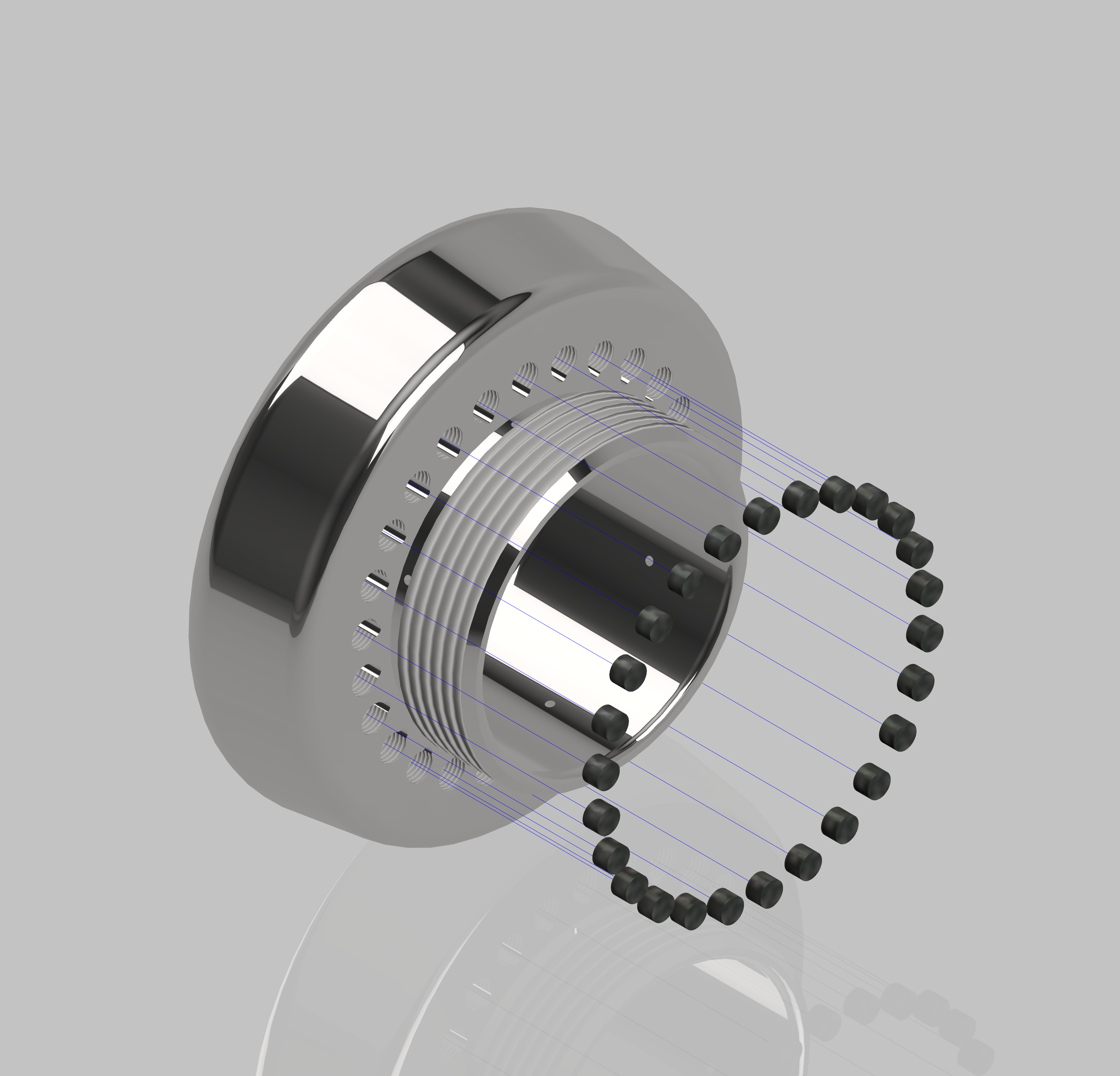}
    \caption{\textbf{Magnet - High Voltage Electrode Assembly} Exploded view of the permanent magnet - high voltage electrode assembly.}
    \label{fig:mag_hv}
\end{figure}

We present a variant of the presented DC accelerator design with embedded permanent magnets. 
The electrode was designed for use in a compact \SI{100}{\kilo\electronvolt} DC-RF photoinjector for time-resolved electron diffraction (currently unpublished work A. Patwardhan et al.).
A simplified exploded view of the magnet-electrode assembly is shown in Figure (\ref{fig:mag_hv}).
In the real implementation of the principle, appropriate precautions were taken to ensure venting of all trapped pockets (to prevent virtual leaks), recessed features to prevent high electric field gradient at corners and edges, and non-magnetic materials of construction. 
The presented design ensures the zero field point with the same magnetic field gradient (\SI{0.2}{\tesla\per\meter}) at the source point.

By mechanically referencing the magnet assembly to the high voltage electrode, the misalignment (as defined by the true position geometric dimensioning tolerance) of the magnet to the high voltage electrode can be better than \SI{20}{\micro\meter} even with routine manufacturing techniques.
This is sufficiently accurate for even the most demanding situations in terms of the apparent emittance growth.

\section{Analysis of Errors and Mitigation Strategies: Extraordinary Performance from Ordinary Components}
\label{sec:analysis_errors}
We continue our analysis of the alignment accuracy of the magnetic field with an analysis of the magnets themselves, i.e. for the following discussion, we assume that the positioning of the magnets is `exact'.
The dimensional accuracy for even off the shelf `fridge magnets' is better than \SI{20}{\micro\meter}\footnote{Measured tolerance from off-the-shelf magnets obtained from supermagnete.nl}.
Thus, we confine our attention to errors of the magnetization vector. 
From earlier work on undulator design, a tolerance of better than \SI{3}{\degree} is expected in the magnetization vector\cite{rakowsky_simple_nodate, chang_magnet_nodate}. 
Regarding scalar errors of the magnetic field, we consider the tolerance grades of Nd magnets, from which we estimate a tolerance of 3\%. 
Given these error estimates, we confine our attention almost solely to magnetization errors i.e. in the following discussion, the geometry of the magnet is considered to be `exact'.

At the outset, we note that the beam dynamics is largely invariant for 3\% errors of magnitude of the magnetic field if the beam energy is adjusted accordingly.
Thus, we explore the angular errors of the magnetization.
Any deviation from the desired magnetization vector creates dipole fields on the beam axis. 
Because the focusing effect of solenoidal fields is quadratic in the magnetic field, and the dipole effects are linear, careful minimization of angular error of magnetization is important for obtaining best performance.
The obvious solution to this problem is to enforce strict specification on the magnetization vector.
Indeed, this is always an option, although there are always economic and practical limits to the achievable level of precision.
Thus, we explore designs through which components with moderate manufacturing tolerance can lead to higher performance, and if (as a last measure), the tolerances are tightened further, the performance can only improve.

\subsection{Enhancing Performance through Statistics}
The standard deviation of the mean is linked to the standard deviation of the distribution and the number of samples as 
\begin{equation}
    \sigma_{\bar{X}} = {\sigma_X}/{\sqrt{N}}
\end{equation}
where $N$ is the number of samples. This implies that if instead of using a single magnet, we compose it from tiling several smaller magnets; any variances in the magnetization vector are averaged out on the symmetry axis. 
The only underlying assumption is that the magnet orientations are sufficiently randomized, which is easily realized in practice. 
For say 25 magnets, a reduction of a factor of 5 is realized in the effective tolerance for the magnetic field on the axis (\SI{3}{\degree} tolerance of the magnets is reduced to \SI{0.6}{\degree} tolerance on the axis by averaging). 
Consequently, the true position of the zero of the magnetic field can be guaranteed to be better than $\sim$\SI{150}{\micro\meter} of the source position with no additional effort.
Measured magnetic field profile of such an assembly and analysis of randomized arrangements can be found in the supplementary material. 

\subsection{Systematic Cancellation of Dipole Errors}
\begin{figure}
    \centering
    \includegraphics[width=0.65\linewidth]{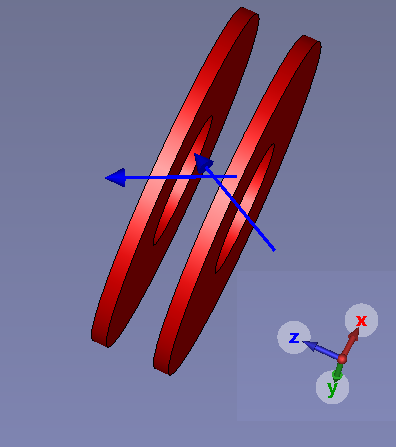}
    \caption{\textbf{Exploded view of the split magnet design:}
    Magnetization vector of the two components of the split magnet is depicted by the arrows. For the purpose of illustration, the tilt with respect to the nominal direction has been exaggerated. Generated with the CST Studio Suite\cite{noauthor_cst_nodate}.
    } 
    \label{fig:splitmagnet}
\end{figure}
\begin{figure}
    \centering
    \includegraphics[width=0.95\linewidth]{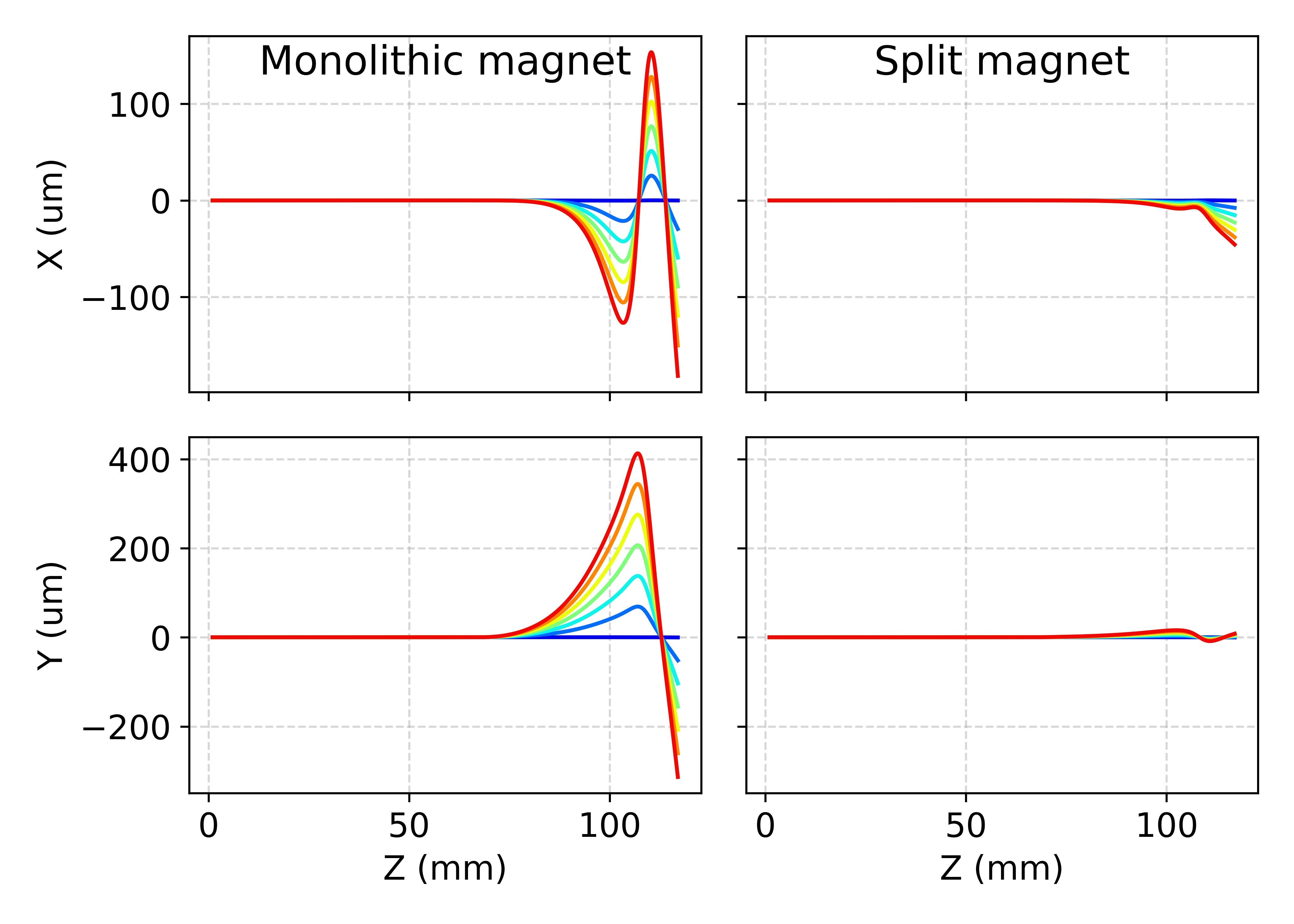}
    \includegraphics[width=0.95\linewidth]{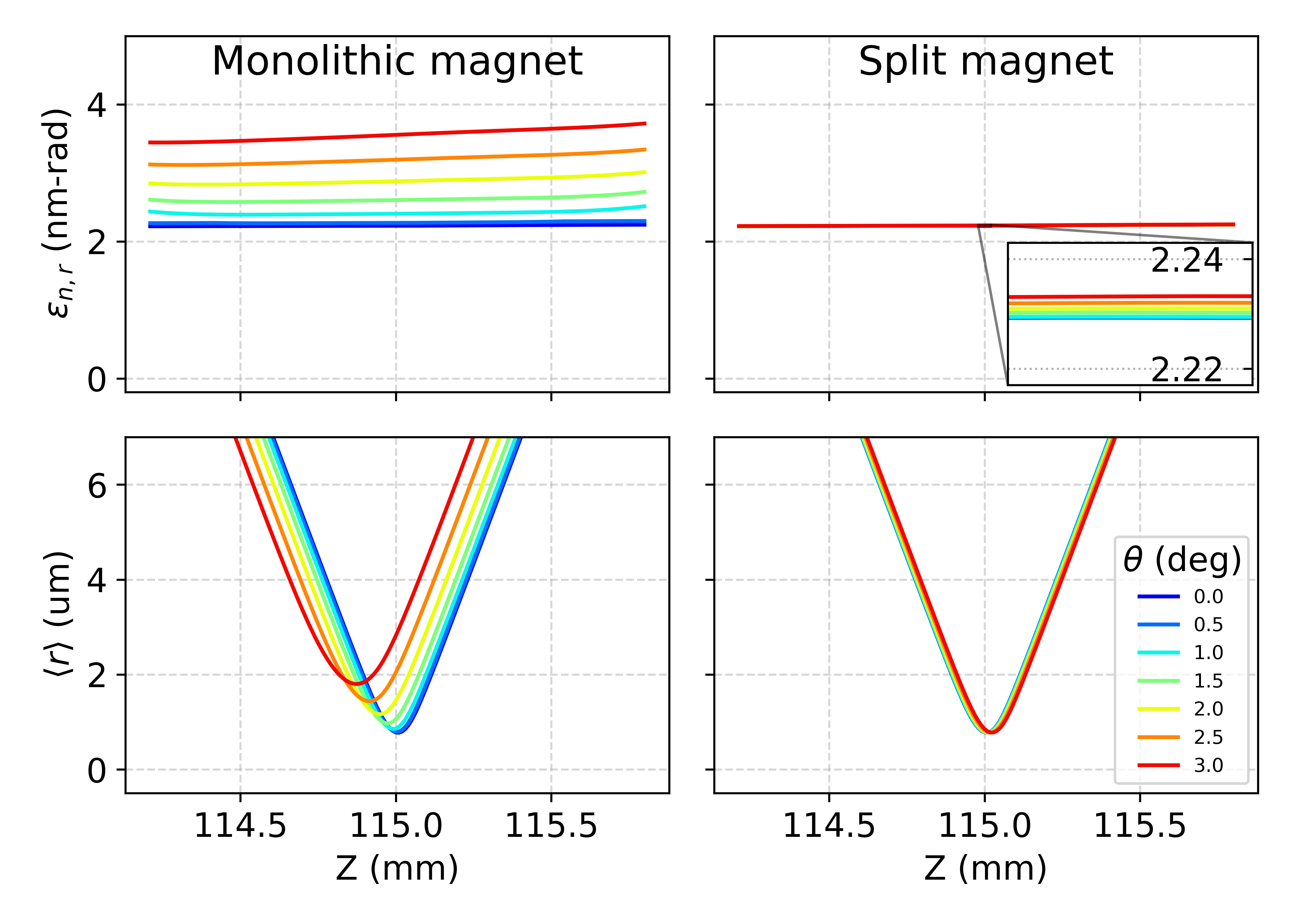}
    \caption{\textbf{Beam Dynamics with Magnetization Errors:}
    \textit{Top:} Beam Deflection due to dipole contribution from magnetization vector error.
    \textit{Bottom:} Effect on Beam Properties due to dipole contribution from magnetization vector error. Only a small fraction of the region near the focus is shown for clarity of illustration. The colors indicate the angular deviation of the magnetization vector with respect to the symmetry axis with red corresponding to \SI{3}{\degree} of misalignment and blue corresponding to perfect alignment.} 
    \label{fig:dipole}
\end{figure}
\vbox{
\begin{quote}
    \textit{``One step forward, two steps back will keep you from running into a wall.''}
    \begin{flushright}{\textit{- Marco Reps\ \  }}\end{flushright}
\end{quote}}

It is worth mentioning that for magnets manufactured in the same `batch', (i.e. magnets that have been magnetized together), are expected to have the same errors. 
This means that the magnetization vector is expected to systematically deviate from the nominal specification. 
For concreteness, we refer to magnet 2 as defined in Section (\ref{sec:design}). 
Magnetization vector error (of say \SI{3}{\degree}) can be considered to be two separate components, an axially magnetized `solenoid' with effective strength reduced by a factor of $\cos\left(3\degree\right) \approx 0.999$ and a dipole with strength corresponding to $\sin\left(3\degree\right) \approx 0.05$. 
The dipole is not uniform along the `z' axis of the beam, but has a profile $B_x (z)$ that can be computed by using an electromagnetic solver.
Because the dipole and solenoid occupy the same section along the beamline, and because the beam dynamics are linear (in $B$)  for dipoles and non-linear (in $B$) for the solenoid, a small vector error of say \SI{3}{\degree} does not necessarily produce a deviation of \SI{3}{\degree}, but can be substantially more especially for a tight focussing geometry. 
When the size of the magnet becomes small (on the order of \SI{1}{\centi\meter}), it is difficult to effectively employ the tiling procedure stated previously.
Instead, we exploit the systematic nature of the errors. 
The magnet is composed of two identical sub-magnets such that the desired magnet configuration is sectioned perpendicular the symmetry axis. Refer Figure (\ref{fig:splitmagnet}). 
One of the two sub-magnets is rotated by \SI{180}{\degree} along the symmetry axis. 
The resulting dipole moments of the two sub-magnets almost entirely cancel out. 
The effect of dipole contributions on both beam position and beam properties can be seen in Figure (\ref{fig:dipole}).
When the length to diameter ratio of the magnets becomes large ($\sim 1$ or greater), the beam rotation around the z-axis changes the optimum rotation of the second magnet with respect to the first.

\section{Conclusion and Outlook}
\label{sec:conclusion}
We have presented an innovative design procedure for a high-performance, micro-focus beamline for low-energy electrons, employing axially magnetized ring-shaped permanent magnets to ensure magnetic field zeros at both source and focus. We have used this procedure to generate a design for the particular case of a beamline based on the ultracold electron source,
In the proposed design, the presence of a magnetic field zero at the source is required, but the approach is generally relevant to low-emittance applications as these magnetic field zeros avoid apparent emittance growth.
Additionally, we propose a number of solutions (self-alignment, statistical averaging, and dipole cancellation) that enable a high-performance magnetic focusing system using simple, off-the-shelf components. 
The proposed techniques and design are likely to provide significant improvements to future low-energy setups in (ultrafast) electron diffraction, microscopy and lithography, as well as DC-RF accelerators.

\section*{Supplementary Information}\label{sec:suppl}
\begin{figure}[ht]
    \centering
    \includegraphics[width=0.95\linewidth]{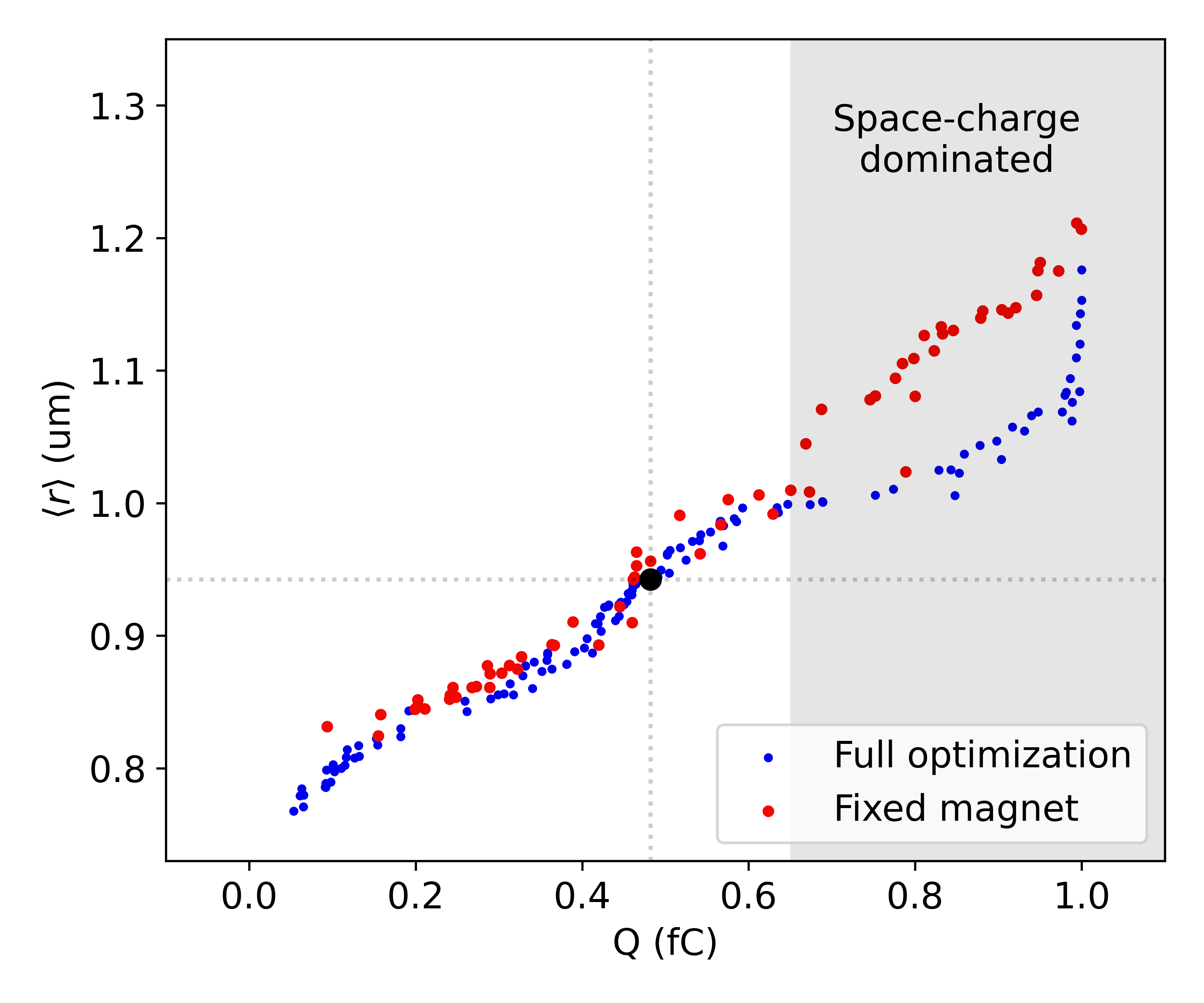}
    \caption{\textbf{Comparison of Pareto fronts} In blue, the pareto front fro the selected focal position (\SI{115}{\milli\meter}) has been reproduced. In red, the accessible pareto front once the magnet configuration has been fixed. The black dot indicates the selected design.}
    \label{fig:ParetoFront_sc}
\end{figure}

\begin{figure}
    \centering
    \includegraphics[width=0.95\linewidth]{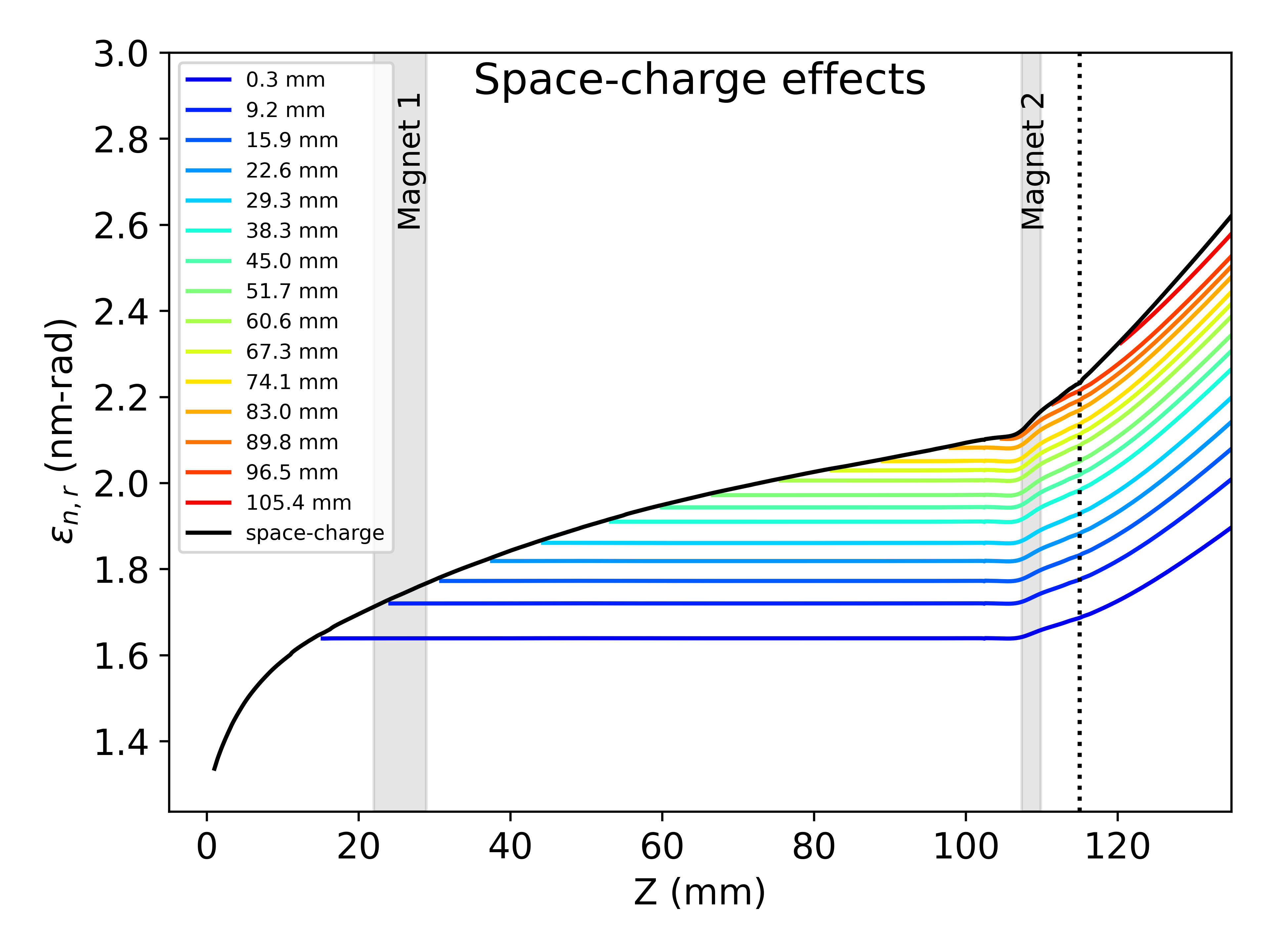}
    \includegraphics[width=0.95\linewidth]{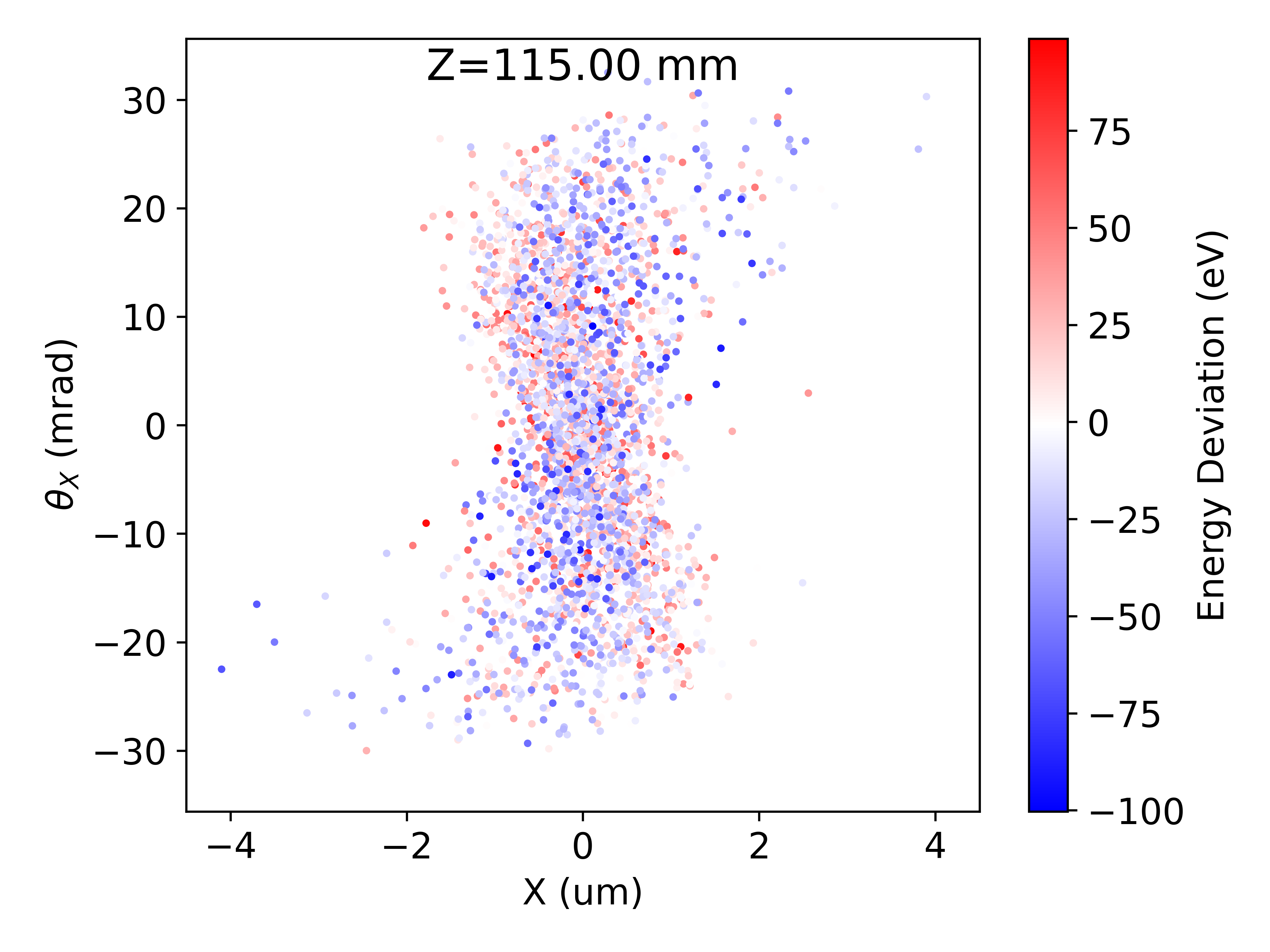}
    \includegraphics[width=0.95\linewidth]{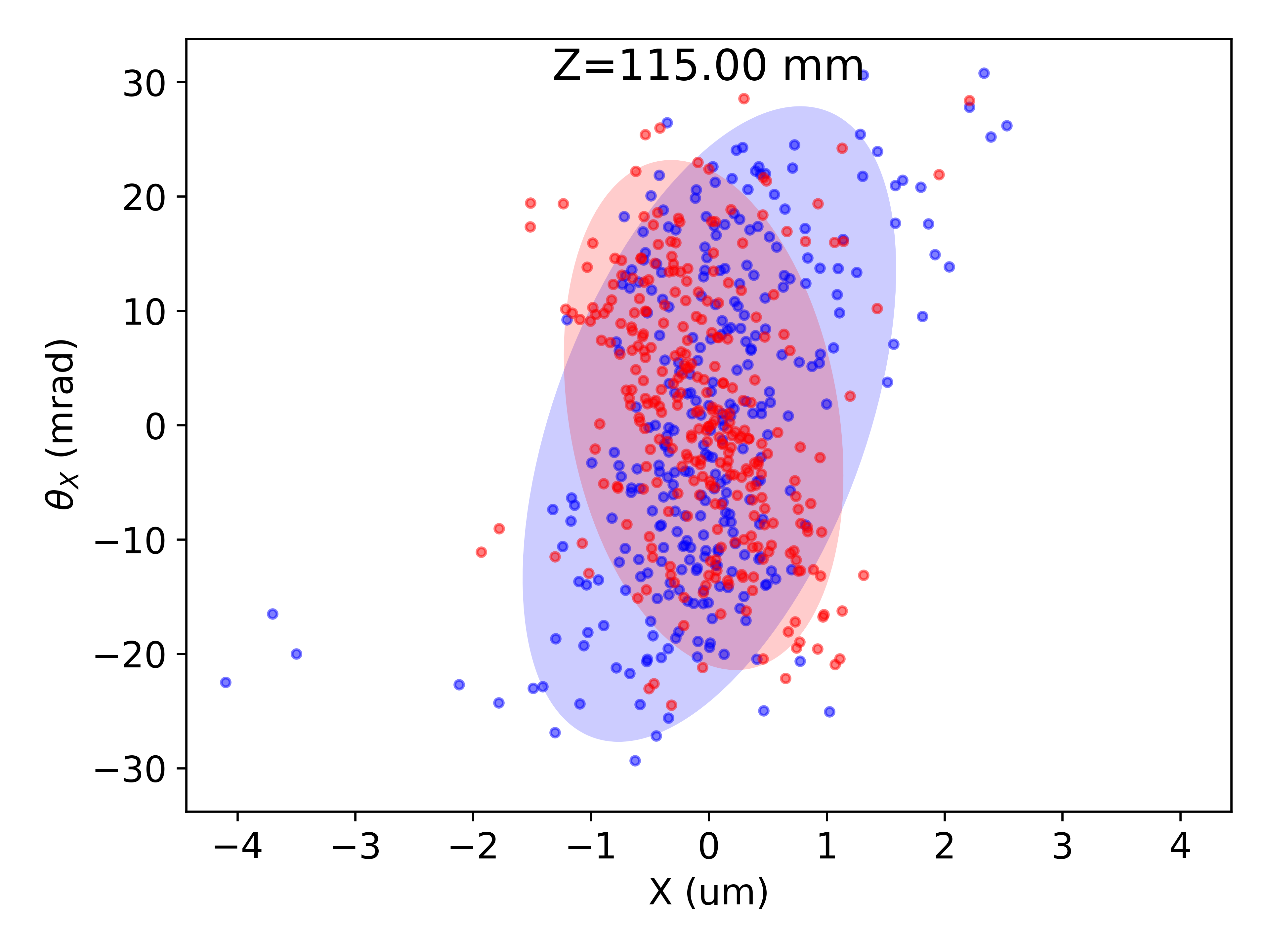}
    \caption{\textbf{Emittance growth from space-charge and chromatic aberrations:} 
    \textit{Top:} Shown in black is the normalized emittance including space-charge effects. In color, are (partial) simulations without space-charge effects. The numbers in the legend correspond to the z-coordinate beyond which the beam is propagated without space-charge effects. 
    \textit{Center:} Transverse Phase Space at the focus, with color of each point corresponding to the energy deviation with respect to the average beam energy. \textit{Bottom:} Transverse Phase Space at the focus for the top and bottom 10\% of electrons in terms of the electron energy. $2\sigma$ confidence ellipeses are plotted for both the top and bottom 10\% of the electrons in terms of energy}
    \label{fig:sc_chromatic}
\end{figure}

\subsection{Generality of the Magnet Design: Space-Charge}
The optimization was constrained to have bunch charges less than \SI{1}{\femto\coulomb} since this represents routine operational parameters of the current iteration of the ultracold electron source.
The optimization of the magnet geometries (and consequently the generation of the Pareto fronts) was performed using a space charge model in which 100 macro particles were used to approximate the total charge. 
This was done to reduce the runtime of simulations. 
Consequently, the numerical values on the $r$ axis in the figures showing Pareto fronts differ slightly from the values quoted in the text. 
The values quoted in the text are from simulations in which case a fine-grained space charge model was used in which each macro particle represents 1 electron. 

With the magnet geometry fixed (Figure (\ref{fig:ParetoFront_f})), the source parameters can still be varied during an experiment. 
Figure (\ref{fig:ParetoFront_sc}) shows in red the Pareto front with the chosen magnet geometry, which will thus be accessible during an experiment. 
A comparison with the Pareto front for the full optimization (in blue) allows for an estimate for the operating range in which space-charge effects are dominant. 
This limit is estimated to be about \SI{0.65}{\femto\coulomb}. 
This bunch charge represents the limit below which changing operational parameters in an experiment does not impact the performance of the system. 
Above this limit, the generality of the design is at the cost of ultimate performance. 

\subsection{Emittance Growth: Space-charge and Chromatic aberration}
In order to understand the effect of space-charge effects and chromatic aberrations, hybrid simulations were performed using the General Particle Tracer.
Simulations were initially performed with full space-charge effects.
At various intermediate times, the simulation was continued without space-charge effects using `initial' particle distribution extracted from the full simulation.
The normalized emittance as computed by the nemirrms routine is plotted in Figure (\ref{fig:sc_chromatic}). 
Space-charge effects are the major contributor to emittance growth until approximately \SI{100}{\milli\meter}.
The linear increase (beyond approximately \SI{100}{\milli\meter}) can be attributed to misaligned slices, which can be verified by plotting slices in the transverse phase.

\subsection{Apparent Emittance: Tolerance Analysis}
We first analyze the emittance growth at the source.
The requirements of atom trapping ensure that the cold atom cloud is located at the point of zero magnetic field. 
The size of this cold atom cloud is approximately \SI{1}{\milli\meter}. 
Thus the maximum axial magnetic field at the source point cannot exceed \SI{0.2}{\milli\tesla} based on the field gradient of \SI{0.2}{\tesla\per\meter}.
For a typical source size $r_0$ of \SI{30}{\micro\meter}, this apparent emittance is about \SI{13}{\pico\meter\radian}, which is much smaller than the intrinsic emittance of the order of \SI{1}{\nano\meter\radian}. 
At the focus of the beam, we do not have a reference dimension to compute the apparent emittance growth; instead, we use the emittance from the particle tracking simulations to estimate the effective tolerance zone in the `z' dimension based on the (known) magnetic field gradient.
The inverted version of Equation (\ref{eq:apparent_emittance}) gives a constraint of $|B_z| <$ \SI{32}{\tesla}, which is an extremely large tolerance window. 
This large tolerance can be attributed to the large convergence/divergence angle of the beam at the focus and correspondingly small focal spot size.
Thus the apparent emittance is not a limiting factor in the presented design, though the constraints still need to be enforced at the souce for atom trapping (and can be desirable at the focus).

\subsection{Experimental Data on Prototype Magnet}
\begin{figure}
    \centering
    \includegraphics[width=0.95\linewidth]{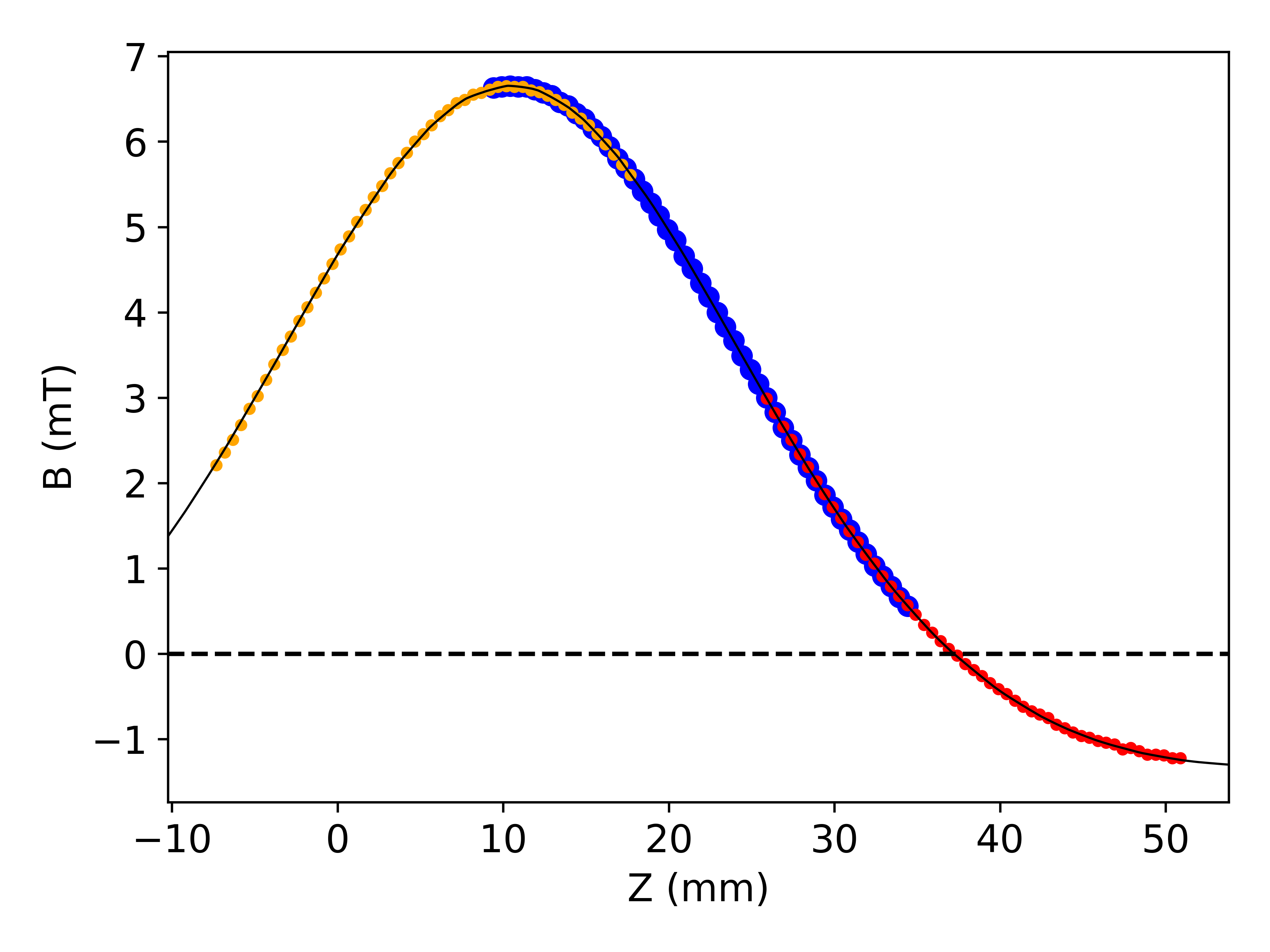}
    \caption{\textbf{Measured Magnetic Field Profile: } On axis field profile of Magnet 1 measured with a Hall effect sensor. The measured data is depicted by red, blue and orange markers. A curve fit based on simulated field profile is depicted in black.}
    \label{fig:measured_profile}
\end{figure}

A prototype of Magnet 1 (discretized design) was manufactured using off the shelf magnets.
The on axis field profile was measured with a Hall effect probe. 
The hall effect probe was first zeroed without the magnet present to zero-out any stray fields. 
The data was fit to a simulated field profile (with fitting parameters related to field scaling, coordinate offsets and background field). 
The results of the measurement and the fit curve are shown in Figure (\ref{fig:measured_profile}). 
The zero-offset from the fit is significantly smaller than the Earth's magnetic field, thus establishing the accurate positioning of the zeros of the magnetic field to a high level of certainty.
The absolute coordinate offset are the offsets from reference datums, which were not controlled during the measurement; and as such are irrelevant to the field profile.
The estimated position of the zero-point has a tolerance zone of around \SI{30}{\micro\meter}.
The field scaling was within 6\% of the simulation result, which is consistent with unknown absolute magnetization of the magnets and calibration uncertainties of the hall effect probe (absolute calibration date unknown). 
The linearity of the probe is assumed (and expected). The errors thereof are expected to be significantly smaller than the absolute calibration errors. 

\subsection{Randomized Evaluation of Uncertainties}
\begin{figure}
    \centering
    \includegraphics[width=0.75\linewidth]{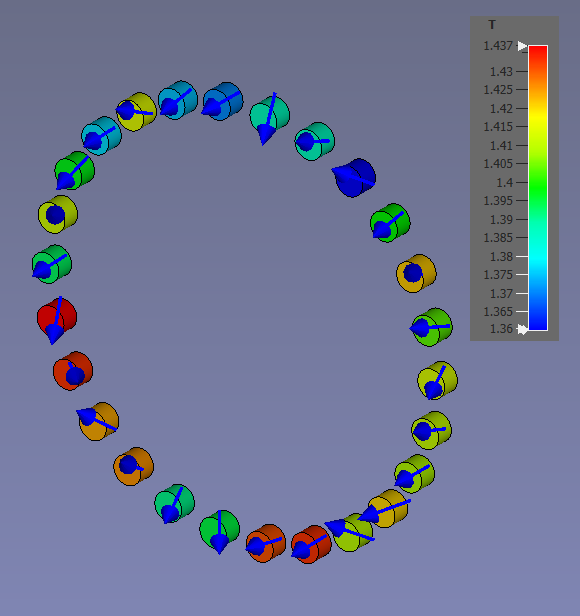}
    \caption{\textbf{Randomized Evaluation of Transverse Fields:} The individual magnetization vectors are depicted by arrows. To enhance clarity of illustration, the orientation error is grossly exaggerated. The colors of the magnet correspond to `scalar' errors, i.e, the magnitude of the magnetization vector. The magnet configuration depicted here corresponds to the HV electrode embedded design.}
    \label{fig:randomized_uncertainty}
\end{figure}

A Monte-Carlo style evaluation of the dipole component of the magnetic field arising from magnetization errors (scalar + vector) of the individual magnets was performed using the CST Studio Suite \cite{noauthor_cst_nodate}.   
The CST VBA scripting interface enables programmatic generation of randomized configurations. 
By evaluating the transverse component of the field, and the `ideal' field gradient, the transverse offset of the zero-field point can be estimated by division.
A picture depicting a randomized configuration is shown in Figure (\ref{fig:randomized_uncertainty}).

\section*{Author Contributions}
A.P.: Conceptualization, Investigation, Formal Analysis, Data Curation, Methodology, Software, Validation, Visualization, Writing – original draft, Writing – review and editing.
B.G.: Methodology, Software, Writing – review and editing.
J.L.: Funding acquisition, Conceptualization, Resources, Writing – review and editing.
J.H.: Funding acquisition, Conceptualization, Supervision, Project administration, Resources, Writing – review and editing.\\\\

\begin{acknowledgments}
The authors acknowledge fruitful discussions with Léon van Velzen (Voltrace).
J.V. Huijts is supported by the European Union’s Horizon 2020 research and innovation programme under the Marie Skłodowska-Curie grant agreement No. 101066850, and by a Branco Weiss Fellowship - Society in Science, administered by the ETH Zürich.
\end{acknowledgments}

\bibliography{ref}

@article{siwick2003atomic,
  title={An atomic-level view of melting using femtosecond electron diffraction},
  author={Siwick, Bradley J and Dwyer, Jason R and Jordan, Robert E and Miller, RJ Dwayne},
  journal={Science},
  volume={302},
  number={5649},
  pages={1382--1385},
  year={2003},
  publisher={American Association for the Advancement of Science}
}

@article{vanoudheusden2010compression,
  title={Compression of subrelativistic space-charge-dominated electron bunches for single-shot femtosecond electron diffraction},
  author={Van Oudheusden, T and Pasmans, PLEM and Van Der Geer, SB and De Loos, MJ and Van Der Wiel, MJ and Luiten, OJ},
  journal={Physical review letters},
  volume={105},
  number={26},
  pages={264801},
  year={2010},
  publisher={APS}
}

@article{lou1998stability,
  title={Stability considerations of permanent magnet quadrupoles for CESR phase-III upgrade},
  author={Lou, W and Hartill, D and Rice, D and Rubin, D and Welch, J},
  journal={Physical Review Special Topics-Accelerators and Beams},
  volume={1},
  number={2},
  pages={022401},
  year={1998},
  publisher={APS}
}

@article{halbach1980design,
  title={Design of permanent multipole magnets with oriented rare earth cobalt material},
  author={Halbach, Klaus},
  journal={Nuclear instruments and methods},
  volume={169},
  number={1},
  pages={1--10},
  year={1980},
  publisher={Elsevier}
}

@Inbook{DebMOO2005,
author="Deb, Kalyanmoy",
editor="Burke, Edmund K.
and Kendall, Graham",
title="Multi-Objective Optimization",
bookTitle="Search Methodologies: Introductory Tutorials in Optimization and Decision Support Techniques",
year="2005",
publisher="Springer US",
address="Boston, MA",
pages="273--316",
isbn="978-0-387-28356-2",
doi="10.1007/0-387-28356-0_10",
url="https://doi.org/10.1007/0-387-28356-0_10"
}

@book{metcalf1999laser,
  title={Laser cooling and trapping},
  author={Metcalf, Harold J and Van der Straten, Peter},
  year={1999},
  publisher={Springer Science \& Business Media}
}

@article{nshii_surface-patterned_2013,
	title = {A surface-patterned chip as a strong source of ultracold atoms for quantum technologies},
	volume = {8},
	copyright = {2013 Springer Nature Limited},
	issn = {1748-3395},
	url = {https://www.nature.com/articles/nnano.2013.47},
	doi = {10.1038/nnano.2013.47},
	number = {5},
	urldate = {2024-04-23},
	journal = {Nature Nanotechnology},
	author = {Nshii, C. C. and Vangeleyn, M. and Cotter, J. P. and Griffin, P. F. and Hinds, E. A. and Ironside, C. N. and See, P. and Sinclair, A. G. and Riis, E. and Arnold, A. S.},
	month = may,
	year = {2013},
	note = {Publisher: Nature Publishing Group},
	pages = {321--324},
}

@article{mcgilligan_grating_2017,
	title = {Grating chips for quantum technologies},
	volume = {7},
	copyright = {2017 The Author(s)},
	issn = {2045-2322},
	url = {https://www.nature.com/articles/s41598-017-00254-0},
	doi = {10.1038/s41598-017-00254-0},
	number = {1},
	urldate = {2024-04-23},
	journal = {Scientific Reports},
	author = {McGilligan, James P. and Griffin, Paul F. and Elvin, Rachel and Ingleby, Stuart J. and Riis, Erling and Arnold, Aidan S.},
	month = mar,
	year = {2017},
	note = {Publisher: Nature Publishing Group},
	pages = {384},
}

@article{cotter_design_2016,
	title = {Design and fabrication of diffractive atom chips for laser cooling and trapping},
	volume = {122},
	issn = {1432-0649},
	url = {https://doi.org/10.1007/s00340-016-6415-y},
	doi = {10.1007/s00340-016-6415-y},
	number = {6},
	urldate = {2024-04-23},
	journal = {Applied Physics B},
	author = {Cotter, J. P. and McGilligan, J. P. and Griffin, P. F. and Rabey, I. M. and Docherty, K. and Riis, E. and Arnold, A. S. and Hinds, E. A.},
	month = jun,
	year = {2016},
	pages = {172},
}

@article{krenn_stark_1997,
	title = {Stark effect investigations of resonance lines of neutral potassium, rubidium, europium and gallium},
	volume = {41},
	issn = {1431-5866},
	url = {https://doi.org/10.1007/s004600050315},
	doi = {10.1007/s004600050315},
	number = {4},
	urldate = {2024-04-23},
	journal = {Zeitschrift für Physik D Atoms, Molecules and Clusters},
	author = {Krenn, C. and Scherf, W. and Khait, O. and Musso, M. and Windholz, L.},
	month = dec,
	year = {1997},
	pages = {229--233},
}

@phdthesis{franssen_ultracold_dissertation_2019,
	address = {Eindhoven},
	type = {Phd {Thesis} 1 ({Research} {TU}/e / {Graduation} {TU}/e)},
	title = {An ultracold and ultrafast electron source},
	school = {Technische Universiteit Eindhoven},
	author = {Franssen, Jim},
	month = may,
	year = {2019},
	note = {ISBN: 9789038647319},
}

@article{franssen_compact_2019,
	title = {Compact ultracold electron source based on a grating magneto-optical trap},
	volume = {22},
	url = {https://link.aps.org/doi/10.1103/PhysRevAccelBeams.22.023401},
	doi = {10.1103/PhysRevAccelBeams.22.023401},
	number = {2},
	urldate = {2024-04-21},
	journal = {Physical Review Accelerators and Beams},
	author = {Franssen, J. G. H. and de Raadt, T. C. H. and van Ninhuijs, M. A. W. and Luiten, O. J.},
	month = feb,
	year = {2019},
	note = {Publisher: American Physical Society},
	pages = {023401},
}

@phdthesis{de_raadt_ultracold_2024,
	address = {Eindhoven},
	type = {Phd {Thesis} 1 ({Research} {TU}/e / {Graduation} {TU}/e)},
	title = {The ultracold electron source: from experiment to instrument},
	shorttitle = {The ultracold electron source},
	school = {Eindhoven University of Technology},
	author = {de Raadt, Tim Christiaan Hendrik},
	month = jan,
	year = {2024},
	note = {ISBN: 9789038659190},
}

@article{de_raadt_subpicosecond_2023,
	title = {Subpicosecond {Ultracold} {Electron} {Source}},
	volume = {130},
	url = {https://link.aps.org/doi/10.1103/PhysRevLett.130.205001},
	doi = {10.1103/PhysRevLett.130.205001},
	number = {20},
	urldate = {2023-11-17},
	journal = {Physical Review Letters},
	author = {de Raadt, T.C.H. and Franssen, J.G.H. and Luiten, O.J.},
	month = may,
	year = {2023},
	note = {Publisher: American Physical Society},
	pages = {205001},
}

@article{claessens_ultracold_2005,
	title = {Ultracold {Electron} {Source}},
	volume = {95},
	url = {https://link.aps.org/doi/10.1103/PhysRevLett.95.164801},
	doi = {10.1103/PhysRevLett.95.164801},
	number = {16},
	urldate = {2024-04-23},
	journal = {Physical Review Letters},
	author = {Claessens, B. J. and van der Geer, S. B. and Taban, G. and Vredenbregt, E. J. D. and Luiten, O. J.},
	month = oct,
	year = {2005},
	note = {Publisher: American Physical Society},
	pages = {164801},
}

@article{claessens_cold_2007,
	title = {Cold electron and ion beams generated from trapped atoms},
	volume = {14},
	issn = {1070-664X},
	url = {https://doi.org/10.1063/1.2771518},
	doi = {10.1063/1.2771518},
	number = {9},
	urldate = {2024-04-23},
	journal = {Physics of Plasmas},
	author = {Claessens, B. J. and Reijnders, M. P. and Taban, G. and Luiten, O. J. and Vredenbregt, E. J. D.},
	month = sep,
	year = {2007},
	pages = {093101},
}

@article{yoon_characteristics_2007,
	title = {Characteristics of single-atom trapping in a magneto-optical trap with a high magnetic-field gradient},
	volume = {80},
	issn = {1742-6596},
	url = {https://iopscience.iop.org/article/10.1088/1742-6596/80/1/012046},
	doi = {10.1088/1742-6596/80/1/012046},
	urldate = {2025-04-04},
	journal = {Journal of Physics: Conference Series},
	author = {Yoon, Seokchan and Choi, Youngwoon and Park, Sangbum and Ji, Wangxi and Lee, Jai-Hyung and An, Kyungwon},
	month = sep,
	year = {2007},
	pages = {012046},
}

@misc{pulsar_physics_general_nodate,
	title = {General {Particle} {Tracer}},
	url = {www.pulsar.nl/gpt},
	author = {{Pulsar Physics}},
    year={2025}
}

@misc{noauthor_voltrace_nodate,
	title = {{Voltrace, https://voltrace.io/}},
    year={2025},
	url = {https://voltrace.io/},
}

@misc{noauthor_cst_nodate,
	title = {{CST Studio Suite}},
    year={2025},
	url = {https://www.3ds.com/products/simulia/cst-studio-suite},
}

@article{luiten2007ultracold,
  title={Ultracold electron sources},
  author={Luiten, OJ and Claessens, BJ and Van Der Geer, SB and Reijnders, MP and Taban, G and Vredenbregt, EJD},
  journal={International Journal of Modern Physics A},
  volume={22},
  number={22},
  pages={3882--3897},
  year={2007},
  publisher={World Scientific}
}

@article{engelen2013high-coherence,
  title={High-coherence electron bunches produced by femtosecond photoionization},
  author={Engelen, WJ and Van Der Heijden, MA and Bakker, DJ and Vredenbregt, EJD and Luiten, OJ},
  journal={Nature communications},
  volume={4},
  number={1},
  pages={1693},
  year={2013},
  publisher={Nature Publishing Group UK London}
}

@article{taban2010ultracold,
  title={Ultracold electron source for single-shot diffraction studies},
  author={Taban, G and Reijnders, MP and Fleskens, B and Van der Geer, SB and Luiten, OJ and Vredenbregt, EJD},
  journal={Europhysics Letters},
  volume={91},
  number={4},
  pages={46004},
  year={2010},
  publisher={IOP Publishing}
}

@article{mcculloch2011arbitrarily,
  title={Arbitrarily shaped high-coherence electron bunches from cold atoms},
  author={McCulloch, AJ and Sheludko, DV and Saliba, SD and Bell, SC and Junker, M and Nugent, KA and Scholten, RE},
  journal={Nature Physics},
  volume={7},
  number={10},
  pages={785--788},
  year={2011},
  publisher={Nature Publishing Group UK London}
}

@article{mcculloch2013high,
  title={High-coherence picosecond electron bunches from cold atoms},
  author={McCulloch, AJ and Sheludko, DV and Junker, M and Scholten, RE},
  journal={Nature communications},
  volume={4},
  number={1},
  pages={1692},
  year={2013},
  publisher={Nature Publishing Group UK London}
}

@book{rao_engineering_2014,
	title = {An {Engineering} {Guide} {To} {Photoinjectors}},
	url = {http://arxiv.org/abs/1403.7539},
	urldate = {2023-10-19},
	publisher = {arXiv},
	author = {Rao, Triveni and Dowell, David H.},
	month = mar,
	year = {2014},
	note = {arXiv:1403.7539 [physics]},
}

@book{egerton2005physical,
  title={Physical principles of electron microscopy},
  author={Egerton, Ray F and others},
  volume={56},
  year={2005},
  publisher={Springer}
}

@article{chen2015nanofabrication,
  title={Nanofabrication by electron beam lithography and its applications: A review},
  author={Chen, Yifang},
  journal={Microelectronic Engineering},
  volume={135},
  pages={57--72},
  year={2015},
  publisher={Elsevier}
}

@book{reiser_theory_2008,
	title = {Theory and {Design} of {Charged} {Particle} {Beams}},
	isbn = {978-3-527-40741-5},
	url = {https://doi.org/10.1002/9783527622047},
	author = {Reiser, Martin},
	year = {2008},
}

@article{rakowsky_simple_nodate,
	title = {A {Simple} {Model}-based {Magnet} {Sorting} {Algorithm} for {Planar} {Hybrid} {Undulators}},
    year={2019},
    journal={Proceedings of 10th Int. Particle Accelerator Conf., Kyoto, Japan},
	author = {Rakowsky, G},
}

@article{jayamanna_single_2018,
	title = {Single {Ring} {Permanent} {Magnet} {Lens}},
	volume = {IPAC2018},
	copyright = {CC 3.0},
	url = {http://jacow.org/ipac2018/doi/JACoW-IPAC2018-TUPMK011.html},
	doi = {10.18429/JACOW-IPAC2018-TUPMK011},
	journal = {Proceedings of the 9th Int. Particle Accelerator Conf.},
	author = {Jayamanna, Keerthi and Baartman, Richard and Bylinskii, Yu. and Corwin, Margaret and Planche, Thomas and Simpson, Ryley},
	collaborator = {Todd (Ed.), Satogata and RW (Ed.), Volker, Schaa},
	year = {2018},
}

@mastersthesis{gehrke_design_2013,
	title = {Design of {Permanent} {Magnetic} {Solenoids} for {REGAE}},
	url = {https://inspirehep.net/files/672d5b770a88268f2b928298ec36eda4},
	urldate = {2024-04-23},
	school = {Universität Hamburg},
	author = {Gehrke, Tim},
	year = {2013},
}

@article{chang_magnet_nodate,
	title = {Magnet {Sorting} {Algorithms} for the {SRRC} {EPU5}.6},
	journal = {Proceedings of IPAC’10},
	author = {Chang, C H and Fan, T C and Hsu, I and Hwang, C S and Wang, Ch},
}

@article{zingery_evaluation_1966,
	title = {Evaluation of {Long}‐{Term} {Magnet} {Stability}},
	volume = {37},
	issn = {0021-8979},
	url = {https://doi.org/10.1063/1.1708352},
	doi = {10.1063/1.1708352},
	number = {3},
	urldate = {2026-03-16},
	journal = {Journal of Applied Physics},
	author = {Zingery, W. L. and Whalley, W. B. and Romberg, E. B. and Wheeler, F. W.},
	month = mar,
	year = {1966},
	pages = {1101--1103},
}

@article{willmott_sls_2024,
	title = {{SLS} 2.0 – {The} {Upgrade} of the {Swiss} {Light} {Source}},
	volume = {37},
	issn = {0894-0886},
	url = {https://doi.org/10.1080/08940886.2024.2312059},
	doi = {10.1080/08940886.2024.2312059},
	number = {1},
	urldate = {2026-03-16},
	journal = {Synchrotron Radiation News},
	publisher = {Taylor \& Francis},
	author = {Willmott, Philip R. and Braun, Hans},
	month = jan,
	year = {2024},
	note = {\_eprint: https://doi.org/10.1080/08940886.2024.2312059},
	pages = {24--32},
}

@article{barna_tunable_2025,
	title = {Tunable, unmountable, permanent-magnet-based accelerator magnet},
	volume = {28},
	url = {https://link.aps.org/doi/10.1103/m58n-mjcb},
	doi = {10.1103/m58n-mjcb},
	number = {7},
	urldate = {2026-03-16},
	journal = {Physical Review Accelerators and Beams},
	publisher = {American Physical Society},
	author = {Barna, Dániel and Anda, Gábor},
	month = jul,
	year = {2025},
	pages = {072401},
}

@article{dong_development_2025,
	title = {Development of a novel tunable gradient permanent quadrupole magnet},
	volume = {28},
	url = {https://link.aps.org/doi/10.1103/cwdp-rpfx},
	doi = {10.1103/cwdp-rpfx},
	number = {12},
	urldate = {2026-03-16},
	journal = {Physical Review Accelerators and Beams},
	publisher = {American Physical Society},
	author = {Dong, Shaoxiang and Yang, Yimin and Zhang, Bingshun and Wang, Yiyue and Feng, Guangyao},
	month = dec,
	year = {2025},
	pages = {122401},

}

@article{sanfilippo_magnets_2024,
	title = {Magnets for the {Upgrade} of the {Swiss} {Light} {Source} at the {Paul} {Scherrer} {Institute}-{Design}, {Production}, {Measurement} {Challenges}},
	volume = {34},
	issn = {1558-2515},
	url = {https://ieeexplore.ieee.org/document/10330056},
	doi = {10.1109/TASC.2023.3335029},
	number = {5},
	urldate = {2026-03-16},
	journal = {IEEE Transactions on Applied Superconductivity},
	author = {Sanfilippo, Stephane and Aiba, Masamitsu and Calzolaio, Ciro and Duda, Michal and Gabard, Alexander and Montenero, Giuseppe and Riccioli, Rebecca and Sidorov, Serguei and Vranković, Vjeran and Zoller, Carolin},
	month = aug,
	year = {2024},
	pages = {1--5},
}

@article{shibata_atomic_2019,
	title = {Atomic resolution electron microscopy in a magnetic field free environment},
	volume = {10},
	copyright = {2019 The Author(s)},
	issn = {2041-1723},
	url = {https://www.nature.com/articles/s41467-019-10281-2},
	doi = {10.1038/s41467-019-10281-2},
	number = {1},
	urldate = {2026-05-12},
	journal = {Nature Communications},
	publisher = {Nature Publishing Group},
	author = {Shibata, N. and Kohno, Y. and Nakamura, A. and Morishita, S. and Seki, T. and Kumamoto, A. and Sawada, H. and Matsumoto, T. and Findlay, S. D. and Ikuhara, Y.},
	month = may,
	year = {2019},
	pages = {2308},
}

@article{tsuno_magnetic-field-free_1983,
	title = {Magnetic-{Field}-{Free} {Objective} {Lens} around a {Specimen} for {Observing} {Fine} {Structure} of {Ferromagnetic} {Materials} in a {Transmission} {Electron} {Microscope}},
	volume = {22},
	issn = {1347-4065},
	url = {https://iopscience.iop.org/article/10.1143/JJAP.22.1041},
	doi = {10.1143/JJAP.22.1041},
	number = {6R},
	urldate = {2026-05-12},
	journal = {Japanese Journal of Applied Physics},
	publisher = {IOP Publishing},
	author = {Tsuno, Katsushige and Taoka, Tadami},
	month = jun,
	year = {1983},
	pages = {1041},
}

@article{kohno_new_2017,
	title = {New {STEM}/{TEM} {Objective} {Lens} for {Atomic} {Resolution} {Lorentz} {Imaging}},
	volume = {23},
	issn = {1431-9276},
	url = {https://doi.org/10.1017/S1431927617002963},
	doi = {10.1017/S1431927617002963},
	number = {S1},
	urldate = {2026-05-12},
	journal = {Microscopy and Microanalysis},
	author = {Kohno, Y and Morishita, S and Shibata, N},
	month = jul,
	year = {2017},
	pages = {456--457},
}

@incollection{petford2005lorentz,
 title={Lorentz microscopy},
 author={Petford-Long, AK and Chapman, JN},
 booktitle={Magnetic microscopy of nanostructures},
 pages={67--86},
 year={2005},
 publisher={Springer}
}
\end{document}